\pdfoutput=1
\documentclass[a4paper,usenatbib]{mnras}
\usepackage{amsmath}	
\usepackage{amssymb}	
\usepackage{txfonts}

\usepackage{graphicx}	
\usepackage{color}
\usepackage{booktabs}
\usepackage{rotating}

\makeatletter

\title[Using MOAO to observe metal-poor stars in and towards the Galactic Centre]{Using the Multi-Object Adaptive Optics demonstrator RAVEN to observe metal-poor stars in and towards the Galactic Centre\thanks{Based [in part] on data collected at Subaru Telescope, which is operated by the National Astronomical Observatory of Japan.}}
\author[M. Lamb et al.]{M. Lamb$^{1,2}$\thanks{E-mail: masen@uvic.ca}, K. Venn$^{1}$, D. Andersen$^{2}$, S. Oya$^{3}$, M. Shetrone$^{4}$, A. Fattahi$^{1}$, 
\newauthor L. Howes$^{5,6}$, M. Asplund$^{5}$, O. Lardi{\`e}re$^{2}$, M. Akiyama$^{7}$, Y. Ono$^{7}$, H. Terada$^{3}$,
\newauthor Y. Hayano$^{3}$, G. Suzuki$^{7}$, C. Blain$^{1}$, K. Jackson$^{8}$, C. Correia$^{9}$,  K. Youakim$^{10}$, 
\newauthor and C. Bradley$^{1}$.\\
$^{1}$Department of Physics and Astronomy, University of Victoria, Victoria, British Columbia, V8W 3P2, Canada\\
$^{2}$NRC Herzberg Institute of Astrophysics, 5071 West Saanich Road, Victoria, BC V9E 2E7, Canada\\
$^{3}$Subaru Telescope, NAOJ, 650 North A'ohoku Place, Hilo, Hawaii 96720, USA\\
$^{4}$Mcdonald Observatory, University of Texas at Austin, HC75 Box 1337-MCD, Fort Davis, TX 79734, USA\\
$^{5}$Research School of Astronomy \& Astrophysics, Australian National University, Cotter Rd., Weston, ACT 2611, Australia\\
$^{6}$Lund Observatory, Department of Astronomy and Theoretical Physics, Lund University, Box 43, SE-22100 Lund, Sweden\\
$^{7}$Astronomical Institute, Tohoku University, 6Ð3 Aramaki, Aoba-ku, Sendai, Japan\\
$^{8}$Division of Engineering and Applied Science, California Institute of Technology, 1200 E. California Boulevard, MC 155-44, Pasadena, California 91125, USA\\
$^{9}$Aix Marseille UniversitŽ, CNRS, LAM (Laboratoire d'Astrophysique de Marseille) UMR 7326, 13388 Marseille, France\\
$^{10}$Leibniz Institute for Astrophysics Potsdam, An der Sternwarte 16, 14482 Potsdam, Germany
}

\begin{document}

\pagerange{\pageref{firstpage}--\pageref{lastpage}} \pubyear{2016}

\maketitle

\label{firstpage}

\begin{abstract}

The chemical abundances for five metal-poor stars in and towards the Galactic bulge have been determined from H-band infrared spectroscopy taken with the RAVEN multi-object adaptive optics science demonstrator and the IRCS spectrograph at the Subaru 8.2-m telescope. Three of these stars are in the Galactic bulge and have metallicities between -$2.1<$ [Fe/H] $<-1.5$, and high [$\alpha$/Fe] $\sim$+0.3, typical of Galactic disk and bulge stars in this metallicity range; [Al/Fe] and [N/Fe] are also high, whereas [C/Fe] $<$ +0.3.    An examination of their orbits suggests that two of these stars may be confined to the Galactic bulge and one is a halo trespasser, though proper motion values used to calculate orbits are quite uncertain.  An additional two stars in the globular cluster M22 show [Fe/H] values consistent to within 1 $\sigma$, although one of these two stars has [Fe/H] = -2.01 $\pm$ 0.09, which is on the low end for this cluster.   The [$\alpha$/Fe] and [Ni/Fe] values differ by 2 $\sigma$, with the most metal-poor star showing significantly higher values for these elements.  M22 is known to show element abundance variations, consistent with a multi-population scenario (i.e. \citealt{Marino2009,Marino2011,Alves-Brito2012}) though our results cannot discriminate this clearly given our abundance uncertainties. This is the first science demonstration of multi-object adaptive optics with high resolution infrared spectroscopy, and we also discuss the feasibility of this technique for use in the upcoming era of 30-m class telescope facilities.

\end{abstract}



\section{Introduction}
\label{sec:intro}
The metallicity distribution function (MDF) of the Milky Way (MW) Galaxy is varied and can differ drastically from one Galactic component to the next.  Prime examples of this are the differences in the mean and metal-poor extension of the MDF of the Galactic halo \citep{Schorck2009,Yong2013} compared to that of the Galactic bulge \citep{Hill2011,Ness2013,Bensby2013,Howes2014}. The metal-poor stars in both components are important and can reveal a wealth of information about these environments, i.e. characterization of the Galaxy at earlier stages in its evolution, and constraints on Population III stellar models and chemical yields \citep{Beers2005,Salvadori2007,Ekstrom2008}. Even though most metal-poor stars are found in the halo, it has been proposed that evidence of the most metal-poor stars (even the First Stars) could in fact be located within the Galactic bulge \citep{Diemand2005,Salvadori2010,Tumlinson2010,Gao2010}.  Furthermore, the bulge exhibits evidence of multiple components \citep{Babusiaux2010,Hill2011,Ness2013}, chemically distinct from one another, thus in this paper we have used a new method to search for and examine metal-poor stars in the Galactic bulge, and investigate their locations and orbits. 

The discovery and characterization of metal-poor stars in the bulge has been approached through a variety of techniques. \citet{Bensby2013} discovered a [Fe/H] = -1.9 dex star based on high resolution spectroscopy from a gravitational microlensing event. \citet{GarciaPerez2013} found two stars at [Fe/H] $\sim$ -2.1 dex from near-infrared (NIR) APOGEE\footnote{APOGEE is an H-band, high-resolution, high signal-to-noise spectroscopic survey of thousands of Milky Way stars, carried out at the Apache point observatory.} spectroscopy. More recently, the EMBLA Survey \citep{Howes2014,Howes2015,Howes2016}, \citet{Casey2015}, and \citet{Koch2016} have used photometric indices as metallicity indicators to identify and spectroscopically observe bulge metal-poor stars at high resolution; as a result they have shown the metal-poor tail end of the bulge MDF extends much further than has previously been observed, showing a sample of 23 stars with -2.3~$<$~[Fe/H]~$<$~-4.0. In addition, the ARGOS survey \citep{Ness2013} have identified $\sim$20 stars with [Fe/H]~$<$~-2 at medium resolution.

There have been two notably unique features regarding bulge metal-poor stars to date. The first is that out of the 23 stars observed by \citet{Howes2015}, \textit{none} of the stars were observed to be significantly enhanced in carbon, which is contrary to that found by \citet{Placco2014} in the halo at these metallicities. Second, is a sample of stars observed in NGC 6522 - a bulge cluster with one of the oldest known ages in the Galaxy - show anomalously high s-process elements given their metallicities ([Fe/H] $\sim$ -1.0), a feature only seen in halo stars with [Fe/H]~$<$~-3.0 (this feature however has been argued to be a false detection, e.g. see \citealt{Ness2014}). Discovering and accounting for these unique features within bulge metal-poor stars is inherently interesting; \citet{Howes2015} and \citet{Chiappini2011} draw links between these unique features and First Star remnants. As such, filling in the metallicity gap between these two works (i.e. -1.0~$<$~[Fe/H]~$<$~-2.3) with a sample of bulge stars with detailed chemical abundances is compelling.

To date, detailed spectroscopic observations of metal-poor bulge stars have been observationally difficult due to (i) the stellar obscuration from dust, (ii) the stellar crowding of the highly dense region, and (iii) the rare nature of metal-poor stars within the entire sample \citep[$\sim$1 in 5000,][]{Ness2013}.  The aforementioned observing techniques (i.e. \citealt{Bensby2013,GarciaPerez2013,Howes2015}, amongst others) hold advantages and disadvantages over one another, with each technique mitigating one or two of the issues, however none capable of simultaneously correcting all three.  In this paper, we employ a new approach that can significantly mitigate all three issues simultaneously; this method uses the new RAVEN multi-object adaptive optics (MOAO) instrument \citep{Lardiere2014,Ono2015} with the Infrared Camera and Spectrograph (IRCS, \citealt{Kobayashi2000}) at the Subaru 8.2-m telescope. 

RAVEN employs the technology of MOAO - a flavour of adaptive optics (AO) that allows for multiple windows of correction over a wide field of regard (i.e. on the order of arcminutes). Each window is provided by a mirror on a `pick-off' arm, which patrols the astronomical field and is placed over scientifically interesting objects. Bright stars within the field of regard act as guide stars (including a laser guide star), which is then used to reconstruct the volume of turbulence above the telescope.

RAVEN demonstrates this technology with 2 pick-off arms on an 8-m class telescope, however there may be many more arms on future extremely large telescope (ELT) instruments. ELTs will also employ MOAO to a greater effect; a larger diameter pupil will in turn have a meta-pupil that does not de-correlate as quickly as an 8-m class telescope as a function of altitude. 
This results in the capability of a larger field of regard for ELTs, which in turn allows for a larger availability of guide stars (and therefore sky coverage). In addition, ELT-MOAO will use many laser guide stars and will only rely on natural guide stars for tip/tilt/focus measurements; these measurements can use the full aperture of the telescope and therefore sensitivity (and sky coverage) is increased for an ELT compared to an 8-m telescope. However, demonstrating MOAO on a telescope such as Subaru is important to gain technological insight and test scientific observing strategies for these future MOAO applications.

The RAVEN instrument operates in the near-infrared, allowing relief from dust obscuration. In addition, AO performance improves with respect to increasing wavelengths, allowing for higher Strehl ratios compared to the optical. Traditional stellar spectroscopy is usually practiced in optical regions, however the near-infrared has shown to be an excellent complementary resource to optical stellar abundances, \citep[e.g.][]{Meszaros2015,Holtzman2015,Cunha2015,Smith2013,Lamb2015}. This wavelength region is spectroscopically accessible to many of the light elements, with more numerous C, N, and O features than the optical, where the continuum is formed in the deepest layers of the stellar atmosphere \citep{Alves-Brito2012}. Furthermore, the near-infrared is an excellent regime to examine Fe features, where typical Fe lines have high excitation potential and are thus less sensitive to variations in Teff (e.g., see \citealt{Alves-Brito2012}). Finally, light elements such as Si are readily available in the infrared, whereas they can be more elusive in the optical (e.g., see \citealt{Lamb2015}).

In this paper, we use the AO instrument RAVEN coupled with infrared spectroscopy to determine stellar abundances. Thus, the goals of this paper are to demonstrate MOAO, discuss its unique observing strategies for future applications, and to observe metal-poor stars in and towards the centre of our Galaxy.

\section{Observations and data reduction}

\subsection{RAVEN technical details}
\label{sec:RAVEN}

\begin{figure}
\centering
  \includegraphics[clip=true, trim = 0 0 0 0,width=0.45\textwidth]{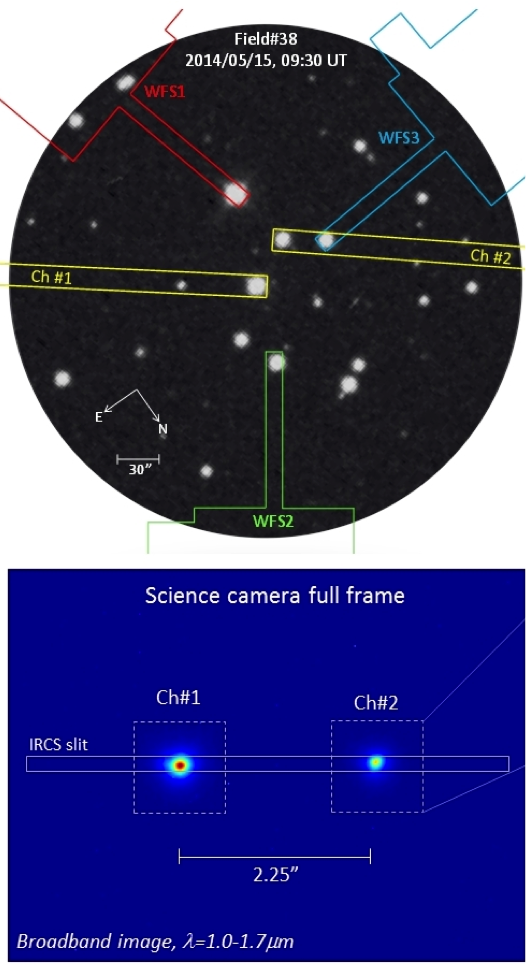}
  \caption{Example configuration of RAVEN's WFS and science channel pick off arms on a field used during an engineering run. The 3 WFS arms are outlined in red, green, and blue while the science channels are outlined in yellow. Also shown is the arrangement of the two channel targets on the IRCS slit.}
  \label{fig:ast}
\end{figure}

RAVEN features two pick off arms, each selecting a 4"x4" region over a field of regard, which can be up to $\sim$ 3.5 arcminutes in diameter. The field of regard is defined by an asterism of three guide stars: either 3 natural or 2 natural and 1 laser (using the existing sodium LGS at the Subaru facility, \citealt{Hayano2010}). The limiting magnitudes on the guide stars are R$\sim$13\footnote{The limiting magnitude could reach R$\sim$14 during dark time, however the moon contaminated they sky during most of our engineering time, and the limiting magnitude depended on its phase and location on the sky}. Each guide star utilizes a wavefront sensor (WFS), from which a tomographic representation of the atmosphere is constructed within the asterism \citep[for more details see][]{Jackson2014,Correia2015}. Any region selected by a pick off arm within the asterism can project this tomographic model onto its own deformable mirror (DM), thus providing an AO correction. The two deformable mirrors are custom 145 actuator ALPAO DMs, with 11 actuators spanning the pupil. MOAO features open-loop control as opposed to the closed-loop control of a classic AO system (otherwise known as SCAO), meaning the system will command the DM using WFS data of the full turbulence as opposed to the residual turbulence. In other words, the WFS does not see the DM correction, because the DM is located after the WFS. However, RAVEN is also capable of employing SCAO and ground layer adaptive optics (GLAO). GLAO is a technique that corrects for ground layer turbulence only, and is employed by averaging the measurements of all three WFSs to drive the DMs. GLAO can be particularly useful when the RAVEN SLODAR showed the majority of the turbulence to be in the ground layer. Figure \ref{fig:ast} shows a typical asterism, defined by the green dashed boundary, as well as two potential science targets within the asterism shown in red.

\begin{figure}
\centering
  \includegraphics[clip=true, trim = 0 0 0 0,width=0.45\textwidth]{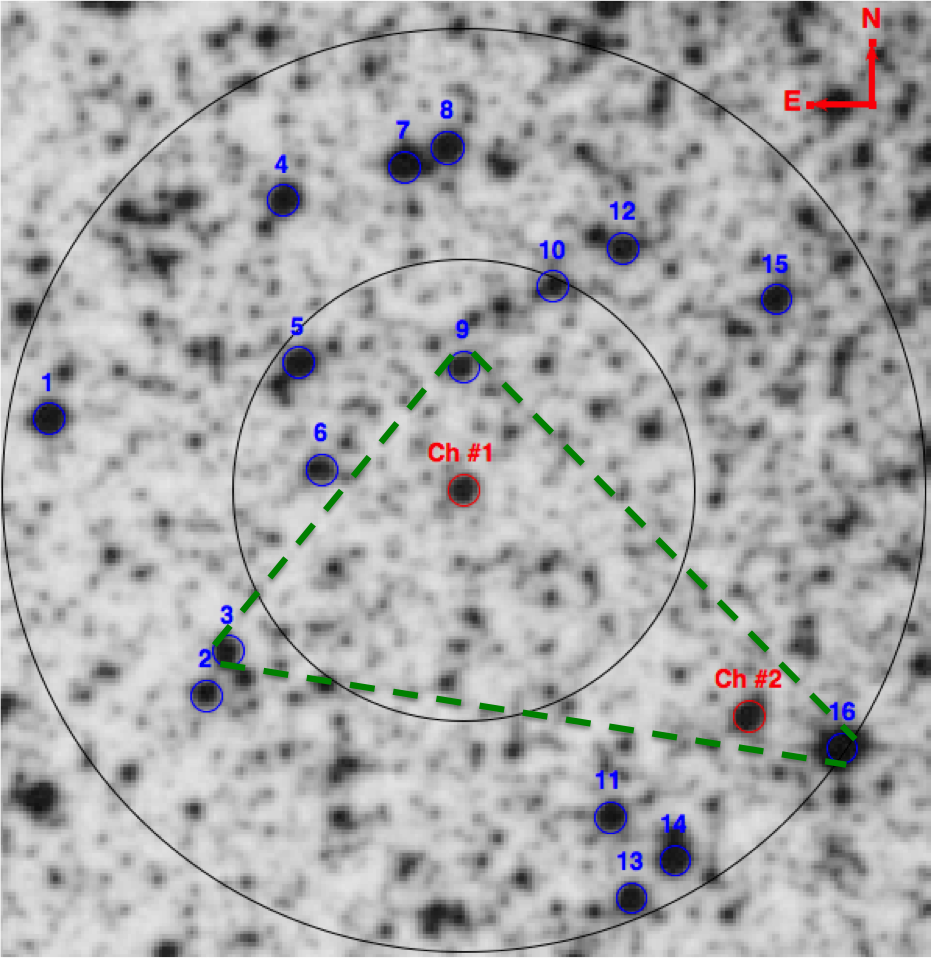}
  \caption{An image of the M22 field used in this work showing the arrangement of suitable guide stars (blue), the adopted asterism (green dashed) and the two science targets (red). The two black circles correspond to 60 and 120 arcseconds. This image was taken from the DSS survey archive (\url{http://archive.eso.org/dss/dss}).}
  \label{fig:ast2}
\end{figure}

RAVEN re-images the two channels onto the slit of the IRCS on the Subaru Telescope.  The IRCS includes two 1024$^2$ ALADDIN III arrays side by side with a wavelength coverage of 0.9-5.6 $\micron$. With one target in each pick off arm, then both stars are directed to opposite sides of the IRCS slit, such that two spectra are gathered simultaneously.  All observations were obtained using the echelle mode of the IRCS with the 0.14 arcsec slit, yielding R $\sim$ 20000 spectra. The H-band wavelength region was chosen (H+ filter\footnote{www.naoj.org/Observing/Instruments/IRCS/echelle/orders.html}), with intermittent coverage from 15000 - 17000 \AA. This region was selected to take advantage of the linelist and techniques for analysing this region developed by the SDSS-III APOGEE project \citep{Smith2013,GarciaPerez2013,Shetrone2015}. 

\subsection{Performance and science observations}

\begin{figure}
\centering
  \includegraphics[clip=true, trim = 530 0 0 0,width=0.75\textwidth]{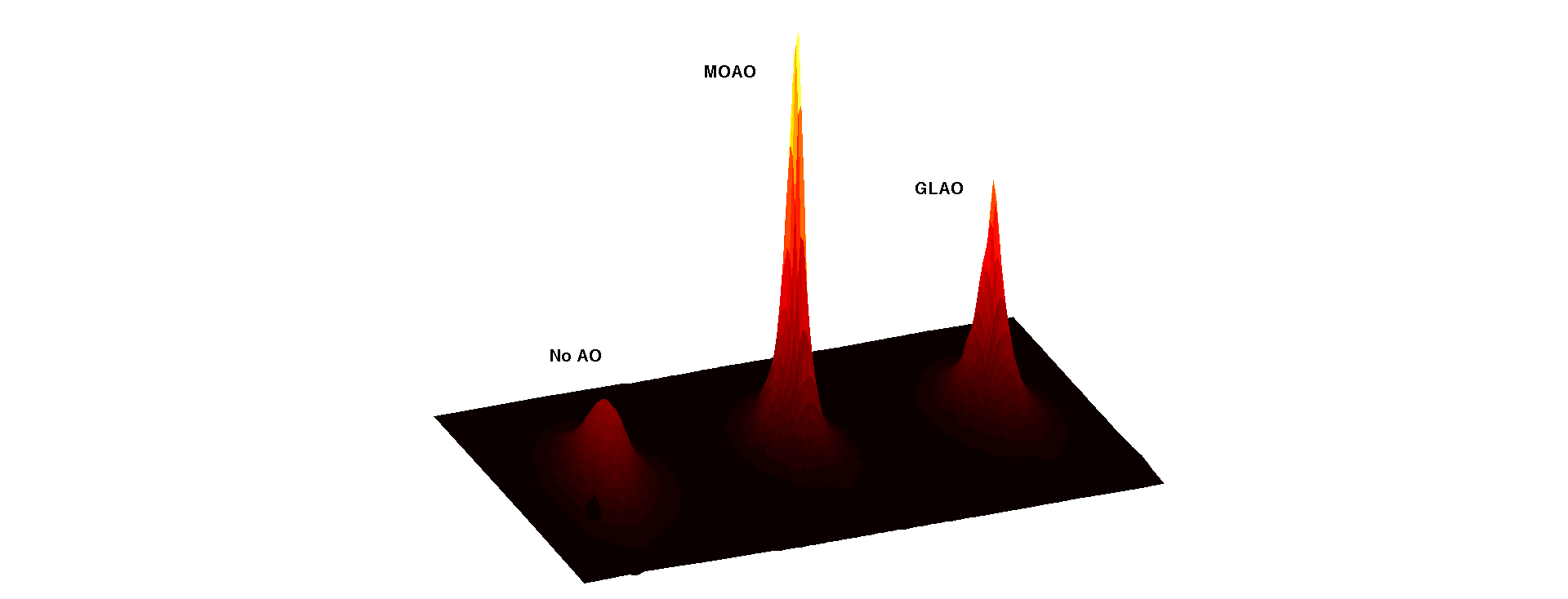}
  \caption{The PSF of MA8 with no AO (left), and with MOAO and GLAO corrections (middle, and right respectively). MOAO is shown to outperform GLAO, however there is substantial improvement from the two corrections. The ensquared energy of the PSFs are 7.29\%, 24.4\%, and 16.19\% for no-AO, MOAO, and GLAO respectively.}
  \label{fig:psfs}
\end{figure}

 Three engineering runs took place between May 2014 and June 2015 at the Subaru Telescope, during which several AO tests were carried out, and the observations of our metal-poor star candidates.  The details of the observations are summarized in Table \ref{table:obs}. Over the four nights where the science observations were taken, the mean seeing of no-AO, SCAO, GLAO, and MOAO was 0.47", 0.09", 0.26", and 0.26" respectively; this shows a general improvement of at least 0.20" in FWHM and demonstrates the AO performance of the instrument. As a result, 2-3 times more flux passed through the slit of our observations, depending on the night and observing mode (MOAO or GLAO). Figure \ref{fig:psfs} demonstrates the quality of the psf for MA8 (prior to incidence on the slit) with no AO correction, and with MOAO and GLAO correction.
 
\subsection {Target selection}
\label{sec:targets}
\subsubsection{Galactic Centre targets}
Three RGB stars were selected for their low metallicity and proximity to the Galactic Centre, based on both SkyMapper photometry and low resolution spectroscopy (from the EMBLA survey, \citealt{Howes2014,Howes2016}). When targeting metal-poor stars in the Galactic Centre, pre-selection is necessary considering that the vast majority of the stars in this region are metal-rich (i.e. \citealt{Hill2011}). Furthermore, observations of these targets in the infrared with AO exploits the issues involved with a crowded and dust-ridden field. These pre-selected targets were observed with RAVEN during two engineering runs (in August 2014, and June 2015). Each pointing involved a simultaneous observation of an additional star of similar brightness to demonstrate the multiplexing capabilities of RAVEN; these additional stars were found to be relatively metal-rich and will be presented in a companion paper. The targets were observed in different AO modes (i.e. MOAO, GLAO, SCAO) depending on the turbulence profile of the evening (further described in Section \ref{sec:GLAO}). 

\subsubsection{M22 targets}
Two metal-poor stars in the globular cluster M22 were selected for our program. This cluster is directly in front of the Galactic bulge but still subject to crowding, with reddening variations of 0.10 mag in \textit{E(B-V)} across the face of the cluster \citep{Marino2011}, thus an ideal target for our MOAO demonstrations.   This cluster also has evidence for a spread in iron, a characteristic found in only a few globular clusters within the Milky Way (i.e. M54, Omega Cen, etc.). It has been argued that M22 is host to multiple stellar populations, with one of the key indicating factors being a significant spread in iron, calcium, A(C+N+O), and s-process elements between stars (e.g. \citealt{Marino2009,Marino2011,Alves-Brito2012}).  However there is also evidence claiming the contrary \citep{Mucciarelli2015}, where it is argued the Fe spread can be explained by the large systematics involved with deriving surface gravities spectroscopically.  
We have selected two targets in M22 from \cite{Lane2011}, who determined Fe abundances for stars in M22 from low resolution spectra (part of the RAVE survey).   Our targets include a metal-poor and a metal-rich RGB star from this sample; the spread in Fe between the two stars is $\Delta \sim$ 0.3 dex. In addition, we required the stars exist within a 3.5' vicinity of each other. Our goal is to determine detailed abundances of the two stars  and search for element abundance differences.  This demonstration can show the advantage of the multiplexing capabilities of RAVEN and the AO correction available in a crowded field, provide homogeneous observations mitigating potential systematic differences in derived abundances, and utilize the infrared to overcome extinction from scattering by dust. An image of the M22 field used in this work, along with a footprint of the asterism, is shown in Figure \ref{fig:ast2}.

\subsubsection{Standard star} 
One standard RGB star in M15 with known low metallicity ([Fe/H] $\sim$ -2.3, \citealt{Meszaros2015}) was observed to track the precision and accuracy of our methods. This star is well described in the literature \citep{Carretta2009,Sobeck2011,Meszaros2015}, providing a comparison sample for all of the the light-element abundances derived in this work. Furthermore, the comparison sample is derived from both optical and infrared spectroscopic methods, providing an excellent framework to determine the validity of our methods. The comparison is drawn in Section \ref{sec:standarAbund}.

\begin{table*}
 \centering
\begin{minipage}{1.0\textwidth}
\centering
  \caption{The sample of stars observed}
  \setlength{\tabcolsep}{2pt}
	\def\arraystretch{1.00}
  \begin{tabular}{@{}lrrccccccccc@{}}
  \hline

  & & & & & & Exp. & & &Obs. &AO &No AO\\\cmidrule(lr){11-11}\cmidrule(lr){12-12}

 2MASS ID (alternate name\footnote{For the rest of this paper we adopt this alternate name when referencing these target stars.})& R.A. (J2000) & Dec. (J2000) & b & l & Date Obs. & (s) & S/N\footnote{Measured at $\sim$ 12-16 different featureless regions across the entire H-band spectrum and reported as a straight average.} &  $V_{hel.}$ (km/s) & mode & \multicolumn{2}{c}{seeing (")\footnote{The median seeing taken for the entire night.}}\\

    \hline
J21295492+1213225 (M15 K341)		&21 29 54.928	&+12 13 22.55  	&-27.27	& 65.05	& Aug. 2014	& 2400	& 80		& -109.59 $\pm$ 1.09		& SCAO\footnote{This mode was used solely to demonstrate the SCAO capabilities of RAVEN and was otherwise unnecessary. A detailed description of the observing modes is in the text.}& 0.06 & 0.46\\
J18364826-2357135 (M22-MA4)		&18 36 48.267	&-23 57 13.56	&-7.66	& 9.89	& June 2015	& 5600	&100		& -142.34 $\pm$ 0.60		& MOAO & 0.36 & 0.53\\
J18364279-2358110 (M22-MA4.1)		&18 36 42.792	&-23 58 11.07	&-7.64	& 9.86	& June 2015	& 5600	&130		& -152.48 $\pm$ 0.59		& MOAO & 0.36 & 0.53\\
J18154190-2749464 (MA8)			&18 15 41.872	&-27 49 45.40 	&-5.20	& 4.24	& Aug. 2014	& 3200	& 80 		& -215.59 $\pm$ 0.98		& MOAO & 0.19 & 0.46\\ 
J16572220-2840402 (MA11)			&16 57 22.207	&-28 40 40.36	&8.85	& -5.79	& June 2015	& 6400	& 110 	& 113.06  $\pm$ 0.45		& GLAO & 0.24 & 0.47\\
J18260509-2536479 (MA14)			&18 26 05.095	&-25 36 47.93	&-6.23	&7.29	& June 2015	& 3600	& 125  	& -136.66 $\pm$ 0.58		& GLAO & 0.24 & 0.41\\

\hline
\label{table:obs}
\end{tabular}
\end{minipage}
\end{table*}

\subsection{Observing strategies with MOAO}

Given that RAVEN observations are made with both a new instrument \textit{and} a new technology, we had to develop unique observing strategies prior to the engineering runs to manage the complexity of MOAO. In some circumstances, it was still necessary to adjust the observation strategy on the fly, while in others we found adjustments were necessary for future observing runs. Since this paper is also intended to provide insight for future MOAO developments, then we discuss these observing strategies here.

\subsubsection{Pick-off arm limitations and arm swapping}
Although the telescope tracks the sidereal rotation, the beam delivered to the Nasmyth focus has a rotating field of view, and was not corrected by Subaru's derotator for several reasons (i.e. degraded image quality). Therefore the field rotation was tracked by the pick-off arms of RAVEN.  As a pick-off arm tracks an object across the field of regard, the optical path length of its respective channel is altered; to conserve this path length, a trombone in the channel translates along the optical axis to compensate for optical path differences created by the varying pick-off arm positions. During the M22 observation, one of the pick-off arms moved beyond the allowable travel of the trombone as it tracked a target throughout the observation. For this reason, we were forced to switch the two targets (and thus pick-off arms) mid observation. Fortunately, this was trivial given that there are two pick off arms, however when an instrument has many pick off arms (such as the TMT IRMOS design), this could be more complicated and something to consider for more complex MOAO systems.

\subsubsection{ABBA nodding}

\label{sec:ABBA}
\begin{figure}
   \centering
    \includegraphics[width=.38\textwidth]{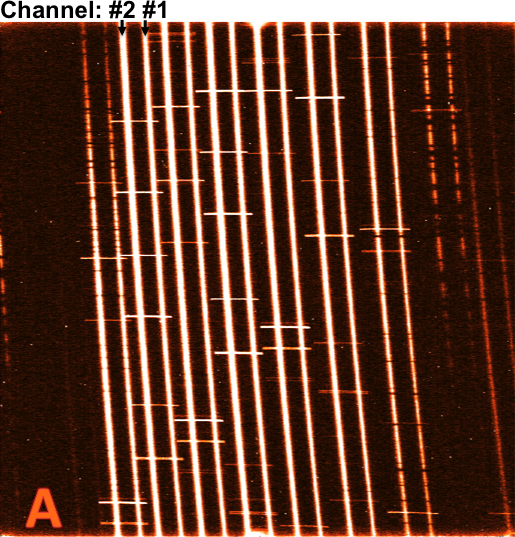}
    \includegraphics[width=.38\textwidth]{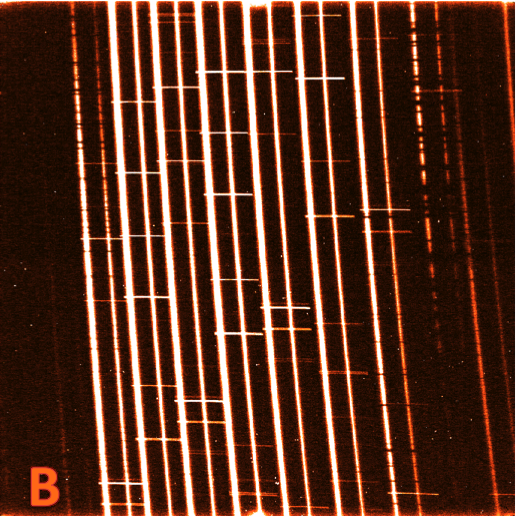}
    \includegraphics[width=.38\textwidth]{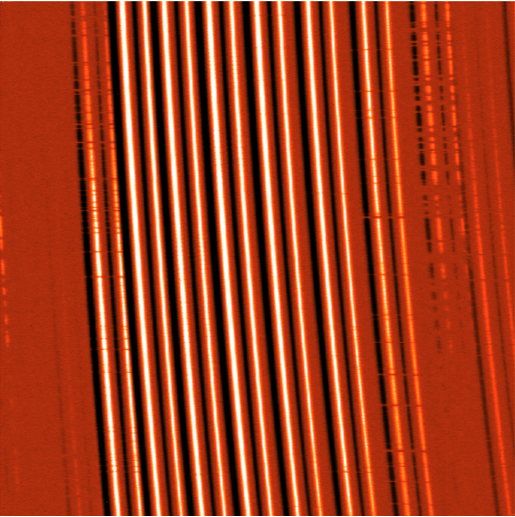}
   \caption{ABBA nodding for M22 MOAO spectra: MOAO allows multiple targets in the cluster to be projected onto the same slit and cross-dispersed side-by-side (labelled as channel \#1 and \#2 on the top-most image above) over several orders. Bottom: Subtracting the top two images from each other yields spectra free of sky lines, dark current and bias; A and B configurations with MOAO were carefully pre-determined to ensure their subtraction did not contain overlapping signals.}
   \label{fig:ABBA}
\end{figure}

To alleviate sky emission lines, dark current and bias effects, spectral data were gathered at two positional configurations projected onto the IRCS slit (referred to as A and B positions), and then subsequent A minus B image pairs are computed. Although this technique is fairly standard for traditional spectroscopic observations, extra care was needed when determining the A and B positions because of the MOAO techniques. For example, typical ABBA nodding involves moving the telescope itself to provide the A and B offset positions. With RAVEN, moving the telescope would knock off a WFS out of its loop, and so the nodding was achieved by moving the pick-off arms instead. When only one star is fed to the spectrograph slit, a subsequent (A - B) image will produce alternating `bright' and `dark' lines corresponding to the positive and negative residual dispersion signals, and the separation between these lines is determined by the size of the nod. However, with RAVEN there are two targets fed to the slit (see the bottom of Figure \ref{fig:ast}), resulting in a much more crowded (A - B) residual dispersion signal image as can be seen by Figure \ref{fig:ABBA} (bottom image). The total slit length is 5.17", and the typical FWHM of the AO-corrected PSFs are 0.2-0.35" (see Table \ref{table:obs}); considering the ABBA nodding should be sufficiently large such that signals do not overlap (i.e. $\sim$1-2") and that \textit{two} PSFs occupy the slit (taking $\sim$2" in total when considering the signal beyond the FWHM), very little slit space is leftover. 

During our first engineering run (May 2015), even after pre-determining suitable A and B nodding configurations, there was slight overlap between signals where the spectra are more closely packed; these spectra were subsequently trimmed where there was signal overlap.   From the experience of our first run we were able to improve on the (A - B) configurations for our future engineering runs.   (A - B) positioning needs careful consideration in more complex MOAO systems, unless such systems send each channel to an independent IFU.

\subsubsection{Choice of GLAO}
\label{sec:GLAO}

Maunakea is known to have most of it's turbulence confined to the ground layer (i.e. \citealt{Chun2009}).  Averaging the WFS measurements from all 3 guide stars during MOAO allowed us to estimate the ground layer turbulence, and we found the performance of GLAO was nearly as good as this averaged MOAO mode when most of the turbulence was observed in the ground layer (through SLODAR measurements). Thus during the nights where we observed MA11 and MA14, there was a significant level of ground layer turbulence and we chose to observe these targets in GLAO mode only.  

\subsubsection{Acquisition}

During our first two engineering runs, we had problems setting the tracking offsets of the ABBA nodding sequence, which resulted in a significant amount of overhead time ($\sim$20-30 min.). Since RAVEN's primary purpose is to demonstrate MOAO our science observations were not optimized and the limited testing of the AO system (once installed on the telescope) was dedicated to engineering items. As a result, the overheads were progressively reduced as we became more familiar with the science observation requirements. By the third engineering run the target acquisition times were reduced to $\sim$10 minutes (similar to standard AO acquisition times on workhorse instruments such as GNIRS on Gemini North). This problem can also be translated to future ELT MOAO systems, where careful foresight will undoubtedly be required to accommodate the increased complexity of more pick off arms when trying to track many simultaneous targets. A field de-rotator may help improve this issue.

\subsubsection{Guide star strategies}

Scientific observations with MOAO are tightly correlated to the availability of guide stars in the field.  The guide star asterisms required careful planning a priori, including back-up asterisms in case any NGSs are unrecognized binary systems.  A back-up asterism was used during our observations of the bulge target MA11, where the LGS proved to be too faint on one particular night (this could have been due to the beam output power, the particular sodium profile that evening, or a combination of factors). The back-up asterism included fainter NGSs (R$>$14.5, which is technically within the limits of RAVEN's capabilities but reduces the AO correction and performance). We further note that during the first two engineering runs, we had difficulty guiding at R$\sim$14.5, however it worked during the third engineering run.  We attribute this to two factors: i) the Moon was not in the sky, as opposed to the first two runs, reducing the background signal on all of the WFSs, and ii) a newly developed centroid correlation technique for the WFSs \citep[see][]{Andersen2014}. The latter point effectively reduces the photon noise by correlating the well-sampled PSF with a reference image. These points summarize the importance of quality (i.e. bright) asterisms, and the necessity of having a back-up asterism (or multiple back-ups), even though this requirement greatly limits the total observable targets on the sky. This can be alleviated if an asterism of LGSs are used. Future ELT MOAO systems will employ multiple LGSs, and thus allow for much greater sky coverage; this is made possible because of the availability of NGSs for Tip/Tilt/Focus correction due to the wide field.

\subsubsection{Telluric standard}
Telluric standard stars are used to divide out the spectral imprint of the atmosphere.  These observations did not require AO, since the targets are bright and isolated.  Nevertheless, \textit{both} MOAO pick off arms were used; one arm for the telluric standard star and the other on a region of sky only.  This observing strategy alleviated the need for ABBA nodding, i.e., traditional sky subtraction.     

\subsection{Data reduction}
\label{sec:reduc}
ABBA nodding (as described in Section \ref{sec:ABBA}) was used to obtain the spectra of our targets.  Equal exposures of the targets were taken at alternating A and B slit positions, producing a set of \textit{N} A/B pairs, where \textit{N} was the number of images required to obtain a final spectrum with signal to noise of $\sim$ 100. 
The A and B positions were selected such that subtraction of one from the other would remove the thermal and sky emission lines, bias, and dark current, but not contain overlapping target signals.  Exposures were taken in 200 second intervals to avoid saturation of the detector. Each image subtraction creates an `A - B' pair (B \textit{subtracted} from A) from which the pairs were then median combined using IRAF\footnote{IRAF (Image Reduction and Analysis Facility) is distributed by the National Optical Astronomy Observatory, which is operated by the Association of Universities for Research in Astronomy, Inc., under cooperative agreement with the National Science Foundation.}. We chose median combining in order to reduce cosmic ray incidents on the images. The median combined image was fixed for bad pixels using a bad pixel mask (located on Subaru's echelle website as \textit{cam\textunderscore badpix.coo}), an example of such an image is shown at the bottom of Figure \ref{fig:ABBA}. Flat fields were created each night with a uniform lamp set to `on' and `off'; the flat `off' images were subtracted from the `on' images and then median combined.  This flat image was normalized using the IRAF task \textit{apnormalize}.  Each science image was divided by the normalized flat and the A spectra were extracted separate from the B spectra. The remaining reduction steps were applied to the A and B spectra separately. 

The wavelength solution was determined from the OH lines and using IRAF's \textit{ecid}. To find accurate OH lines covering the full spectral range of each aperture, an observation of the sky was used (one of our observations had a misplaced slit alignment where \textit{no} stars fell on the slit, thus it contained only sky lines such as OH). An OH atlas \citep{Rousselot2000} was used, and the sky spectrum was fit with a high-order Chebyshev polynomial (x=5, y=3).  The OH atlas reports spectral lines with vacuum wavelengths, which we shifted to air wavelengths for the remainder of the analysis. In certain cases the OH emission lines were unresolved doublets, and for these we adopted the brighter of the two lines.

Telluric subtraction was done with the spectrum of a hot star (late B-type or early A-type) with broad spectral lines (\textit{vsini} $>$100 km/s) observed at roughly the same airmass as our targets (within 0.1 air masses).  The specific stars used are listed in Table \ref{table:telluric}. 

\begin{figure*}
\begin{minipage}{160mm}
   \centering
    \includegraphics[width=1.05\textwidth]{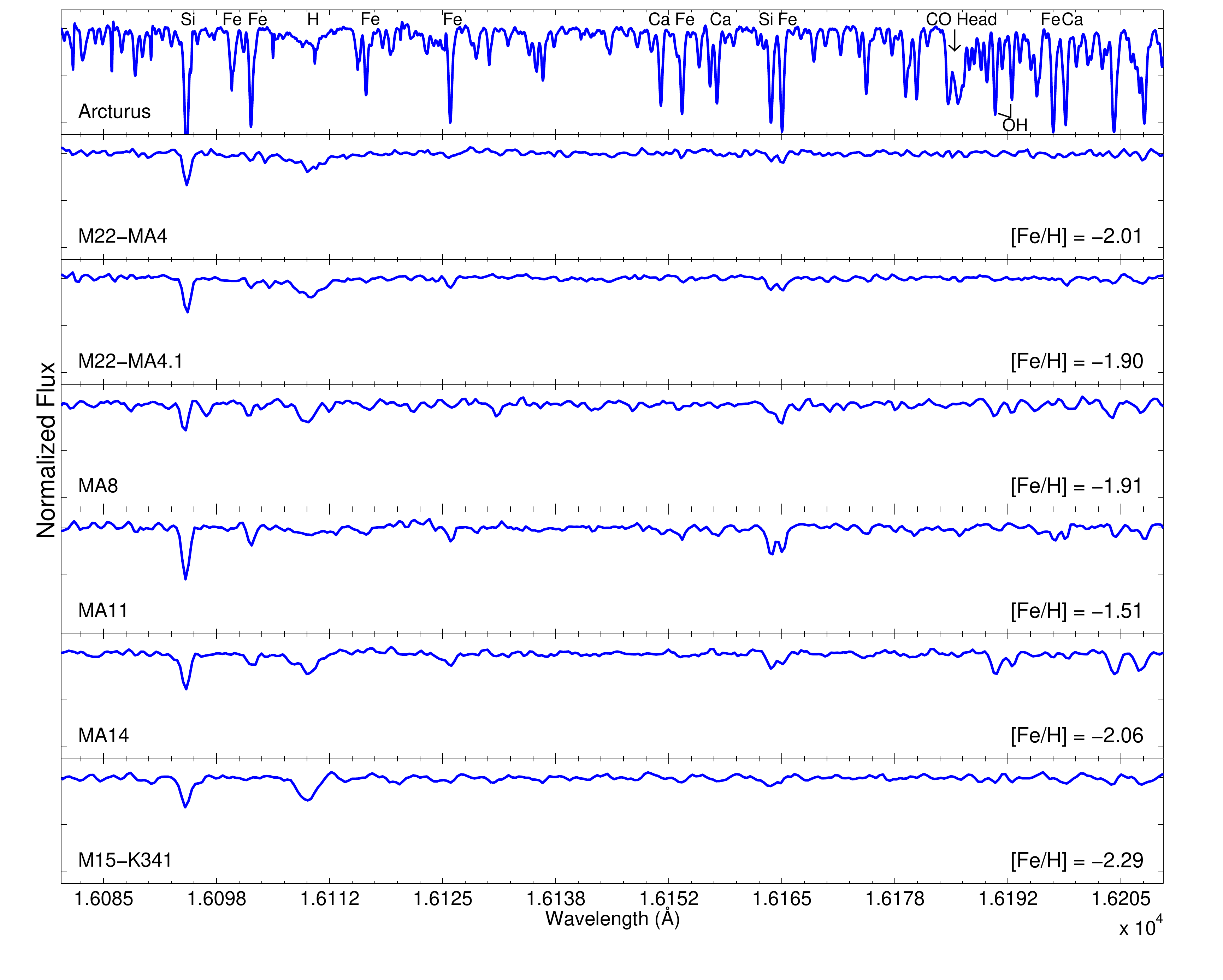}
   \caption{Sample spectral regions Si and Fe lines that were used in the abundance analysis. The derivation of the Fe abundances shown here are discussed in Section \ref{sec:abAnal}. Also shown is the higher-resolution spectrum of Arcturus for comparison purposes only, taken from \protect\citet{HinkleWallace2005}.}
   \label{fig:spectra}
   \end{minipage}
\end{figure*}

Radial velocities were determined from a cross-correlating with the H-band, zero-shifted spectrum of Arcturus \citep{HinkleWallace2005}.  The final A and B spectra were combined to increase the total signal-to-noise ratio. Figure \ref{fig:spectra} shows all of the metal-poor spectra plotted in a region with multiple spectral features used in this analysis.

\begin{table}
\centering
\setlength{\tabcolsep}{2pt}
	\def\arraystretch{1.00}
\caption{Telluric standards}
\begin{tabular}{@{}lcccc@{}}
\hline
Star 		& RA 		& Dec 		& Star(s) Corrected\\
\hline
Sag 4		&17 59 47.55	&-23 48 58.09	& MA 11, MA 8, MA 11\\
HR 7355	&19 24 30.18	&-27 51 57.40	& M22-MA 4, M22-MA 4.1 \\
HD 192425&20 14 16.61	&+15 11 51.39	& M15K341\\

\hline \\[-1.0cm]
\label{table:telluric}
\end{tabular}
\end{table}

\section{Model atmospheres analysis}
\label{sec:modAtm}

Spherically-symmetric, LTE model atmospheres generated by OSMARCS  (\citealt{Gustafsson1975,Gustafsson2008}, also see \citealt{Meszaros2012}) are adopted.
These models are used with the LTE radiative transfer code MOOG (\citealt{Sneden1973}), 
along with the DR10 APOGEE line list \citep{Shetrone2015}.  This abundance analysis follows that in our previous work \citep{Lamb2015}.

\subsection{Stellar parameters}

\label{sec:stellarParams}
The stellar parameters (T$_{\mathrm{eff}}$, log g) of the standard star in M15 were determined through the infrared flux method (IRFM) using calibrations by \citet{Ramirez2005} along with standard stellar structure equations. This method requires optical and infrared photometry of the star, as well as the distance modulus, reddening, and metallicity of the cluster. Microturbulence was determined following the methodology of \citet{Gratton1996} for K-type stars. A consistency check shows that the derived M15 stellar parameters agree well with those of the literature (see Table \ref{table:stellarParams}).  Stellar parameters for the two M22 targets were also computed with the IRFM.  A discrepancy in log(g) of $\sim$ 1 dex was found for M22-MA4 compared to \citet{Lane2011} (RAVE pipeline), however our gravity value is in good agreement with that from the EMBLA survey \citep{Howes2014,Howes2016}.  For the three Galactic bulge targets, we adopt stellar parameters from the EMBLA survey, which infers stellar parameters from the tool known as \textit{sick} (The Spectroscopic Inference Crank; \citealt{Casey2016}). This tool minimizes over all parameters using a Markov chain Monte Carlo approach, sampling the posterior probability distributions to yield best fit stellar parameters along with their uncertainties. We note that for M22-MA4, the parameters from the EMBLA survey and our IRFM results were in good agreement (although this would not take into account significant reddening effects for the bulge stars).  Table \ref{table:stellarParams} summarizes the final stellar parameters for all stars used in this paper.

\begin{table*}
 \centering
 \begin{minipage}{150mm}
 \setlength{\tabcolsep}{2.5pt}
	\def\arraystretch{1.20}
  \caption{Photometry and stellar parameters}
  \begin{tabular}{@{}lcccccccccc@{}}
  \hline
& \multicolumn{7}{c}{Photometry \footnote{B, V, I photometry taken from the sources described in the Ref. column. J, H, and K photometry taken from 2MASS survey.}} & \multicolumn{2}{c}{Cluster parameters\footnote{Taken from \citet{Harris}.}} & Metallicity estimate \footnote{All metallicities estimates are from the EMBLA survey, except for M15 K341 (metallicity taken from \citealt{Carretta2009}) and MA4.1 (\citealt{Lane2011} provide [m/H], which we scale by the MA4 difference: [Fe/H]$_{\mathrm{EMBLA}}$ - [m/H]$_{\mathrm{Lane2011}}$).}\\
\cmidrule(r){2-8}\cmidrule(r){9-10} \cmidrule(r){11-11} 
 Star  & B & V & I & Ref. (BVI)\footnote{A - \citet{Kirby2008}, B - \citet{Libralato2014}, C - \citet{Girard2011}, D - \citet{Monet2003}.} & J & H & K & $(m-M)_V$	& $E(B-V)$ 	& {[Fe/H] (dex)} \\
    \hline
M15 K341		&13.72	&12.54	&11.22	&A	&10.455	&9.796	&9.695	&15.39		&0.10		&  -2.53	\\
M22-MA4			&14.83	&13.60	&12.09	&B	&11.107	&10.514	&10.363	&13.60		&0.34		&  -2.46	\\
M22-MA4.1		&14.38	&13.01	&11.46	&B	&10.467	&9.803	&9.644	&13.60		&0.34		&  -2.26	\\
MA8			 	&14.42	&13.12	&12.16	&C,D&10.576	&9.910	&9.711	& -			& -			&  -2.28	\\
MA11		 	&14.13	&13.12	&11.98	&C,D&11.246	&10.75	&10.658	& -			& -			&  -2.41	\\
MA14		 	&14.99	&13.35	&10.84	&C,D&10.155	&9.387	&9.181	& -			& -			&  -2.35	\\

\hline
 \label{table:phot}
\end{tabular}
\end{minipage}
\end{table*}

\begin{table*}
 \centering
 \hskip-0.2cm
 \begin{minipage}{120mm}
\setlength{\tabcolsep}{2.5pt}
	\def\arraystretch{1.10}
  \caption{Stellar Parameters}
  \begin{tabular}{@{}lcccccccc@{}}
  \hline
			&\multicolumn{2}{c}{EMBLA\footnote{Stellar parameters obtained from the EMBLA Survey \citep{Howes2014,Howes2016}}}&\multicolumn{3}{c}{Other sources\footnote{A - \citet{Carretta2009}, B - \citet{Lane2011}}} &\multicolumn{2}{c}{This study\footnote{Calculated following \citet{Ramirez2005}; photometry taken from Table \ref{table:phot} and errors are discussed in Sec \ref{sec:paramErrors}}}\\
\cmidrule(r){2-3}\cmidrule(r){4-6} \cmidrule(r){7-8} 
Star			& T$_{\mathrm{eff}}$ (K)	& log(g)  & T$_{\mathrm{eff}}$  (K)& log(g)	& 	& T$_{\mathrm{eff}}$ (K)& log(g)  \\
    \hline
M15 K341	&-		        &-		        &4324  $\pm$ 50	    &0.69  $\pm$ 0.20 	& A	&4282$\pm$ 150	&0.52 $\pm$ 0.20\\
M22 MA4	 	&4756 $\pm$ 124  &1.68$\pm$ 0.34	&4514  $\pm$ 276.5  &0.26  $\pm$ 0.60	& B	&4685$\pm$ 150	&1.47 $\pm$ 0.20\\
M22 MA4.1 	&-		        &-		        &4646  $\pm$ 276.5	&0.99  $\pm$ 0.60	& B	&4577$\pm$ 150	&1.18 $\pm$ 0.20\\
MA8		 	&4495 $\pm$ 127 &0.84 $\pm$ 0.35	&-	   	            &-		& -	&-		&-\\
MA11	 	&4514 $\pm$ 122 &0.31 $\pm$ 0.30	&-	    	        &-		& -	&-		&-\\
MA14	 	&4267 $\pm$ 125 &0.74 $\pm$ 0.32	&-	     	        &-		& -	&-		&-\\

\hline
 \label{table:stellarParams}
\end{tabular}
\end{minipage}
\end{table*}

\subsection{Stellar parameter uncertainties}
\label{sec:paramErrors}

Stellar parameter uncertainties for our M15 and M22 targets were determined by altering the input variables associated with the IRFM (reddening, distance modulus to cluster, stellar mass, photometry and metallicity). Results for one star, M22-MA4.1 are shown in Table 6, and adopted for the rest of the M15 and M22 targets.  The reddening error was taken as the total reddening variation (0.10 mag) across the face of M22 from \citet{Marino2011}, an error in the distance modulus of 0.2 mag is taken from \citet{Monaco2004}, the stellar mass uncertainty was taken as 3$\%$, and the metallicity uncertainty as 0.10 dex, following \citet{Lamb2015}. Photometric errors are typically 0.2 mag (following errors reported in the 2MASS survey), however these had negligible impact on the stellar parameter determinations. The mean uncertainties, $\Delta$T$_{\mathrm{eff}}$ and $\Delta$log(g), were determined to be $\pm$ 123 K and $\pm$ 0.15 respectively; we conservatively adopt $\Delta$T$_{\mathrm{eff}}$ = $\pm$ 150 K and $\Delta$log(g) = $\pm$ 0.20 dex to include systematic uncertainties in the calibrations themselves. We adopt an uncertainty in microturbulence of $\Delta\xi$ = 0.5 km/s. For the Galactic bulge targets, we adopt the uncertainties from the EMBLA survey \citep{Howes2014,Howes2016}. The survey inferred stellar parameter uncertainties calculated with the tool \textit{sick}, as discussed in Section \ref{sec:stellarParams}. All stellar parameter uncertainties are presented in Table \ref{table:stellarParams}.

\section{Abundance analysis}
\label{sec:abAnal}

Elemental abundances are determined using the LTE 1D radiative transfer code MOOG\footnote{MOOG was originally written by Chris Sneden (1973), and has been updated and maintained (see Sneden et al. 2012), with the current versions available at http://www.as.utexas.edu/~chris/moog.html.}. The first step in the infrared spectral analysis of a red giant is to determine its Fe and CNO abundances, and potentially refine the model atmosphere adopted. 

Iron has been determined from the spectrum syntheses of 23-50 relatively isolated Fe I lines (see Appendix A). Measurement errors were estimated from each synthetic fit by visual inspection. Spectral broadening parameters were determined from the instrumental resolution, where we adopted Gaussian broadening with FWHM = 0.85 \AA, and made slight adjustments to the local continuum (typically $<$ 1\%) for the best fits. The mean Fe abundance is determined, and its measurement error is the weighted mean of the individual measurements. This measurement error technique is applied for every element and throughout this paper. 

The C and O abundances were determined by following the procedures in \citet{Lamb2015} and \citet{Smith2013}. C was determined from CO features (which are weakly sensitive to O), mainly from features in the band heads starting at 15578, 15775, 15978, 16185, 16398, 16613, and 16835 \AA. In most cases C features at these metallicities are difficult to discern from the continuum, and while upper errors are not difficult to determine the lower errors in some cases are not possible. Therefore we adopt upper limits for C; however to determine O and N, a C value must still be adopted (since OH and CN features depend on C) and so we set the C abundance to its upper limit value. 

O was determined (using the same method as Fe) from a large sample of OH lines; typically 30-40 features which spanned the entire spectrum and had no CN contamination. With the new O abundance, C was re-determined from the CO features, although our stars are sufficiently metal-poor that the CO features are weak.  With this new C abundance (upper limit), then O was re-calculated from the OH lines. This procedure usually required only 1 iteration before the C and O abundances converged. Nitrogen was determined from the CN features between 15100-16100 \AA, although the metal-poor stars in this analysis have relatively weak CN features.

To test the impact of adopting the upper limit value of C on the derived O and N abundances, we perform a simple analysis. Using MA14 (chosen arbitrarily), we decrease the upper limit C abundance by it's \textit{upper error} (0.25 dex); as previously stated the upper error can easily be determined while the lower error is extremely difficult. We re-compute the O and N abundances using this shifted C value and we find O is unchanged while N is \textit{increased} by 0.22 dex (approximately the value of how much C was decreased). We therefore caution the values of our N abundances may be higher than what is reported here, and for this reason our final N abundances are derived as lower limits. We note however, any [(C+N)/Fe] or [(C+N+O)/Fe] calculations should remain relatively unchanged, regardless of these upper and lower limit variations.

\begin{figure*}
    \includegraphics[clip=true,trim = 0 0 0 0,width=0.8\textwidth]{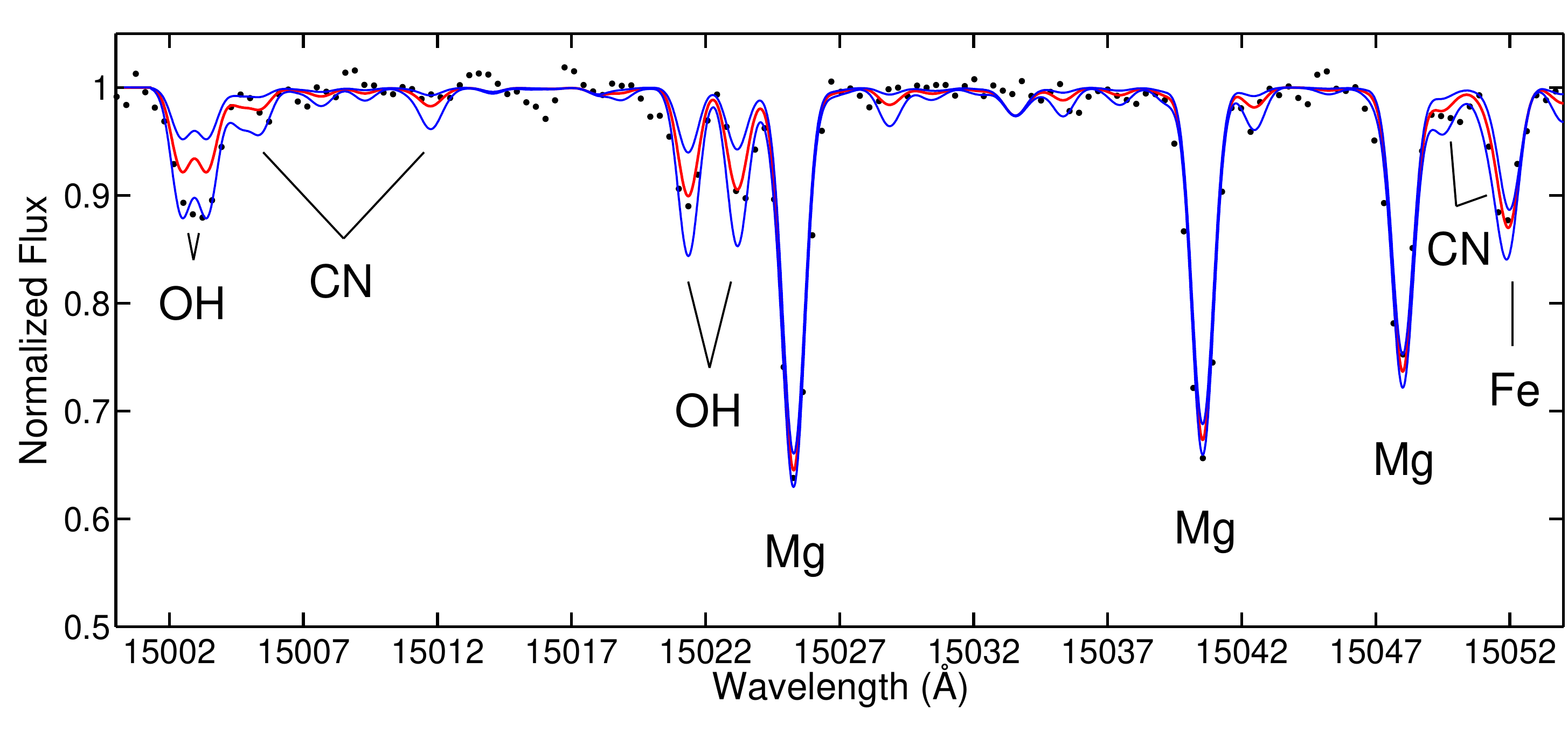}
   \caption{Sample spectral region for MA14; observed spectra are shown as black data points. Also shown is the synthetic spectra computed with the final adopted abundances for this star (solid red line). The syntheses from shifting each element by its adopted upper and lower error is also plotted for reference (shown in blue); the description of how these errors are computed is in Section \ref{sec:abAnal}. For C, the upper error is adopted as the lower error as well (also see Section \ref{sec:abAnal}) for visual purposes.}
   \label{fig:sampleReg}
\end{figure*}

New model atmospheres were interpolated from the OSMARCS atmosphere grid given the new CNO and Fe abudances for each star and the analysis repeated. With the Fe, C, N, and O fixed, the abundances of the remaining elements were determined from individual line syntheses (summarized in Appendix A, and discussed in Section \ref{sec:abund}). A spectral region of MA14 is shown in Figure \ref{fig:sampleReg} along with a best fit spectrum synthesis, and $\pm 1\sigma$ total errors (in the case of C we use its upper error for both upper/lower errors). The abundance errors due to the stellar parameter uncertainties are reported in Table \ref{table:abund}.

\subsection{Standard star comparison}
\label{sec:standarAbund}

A red giant in the metal-poor globular cluster M15 was observed and analyzed as a standard star. This star has been analyzed by the APOGEE survey \cite{Meszaros2015}, therefore a similar analysis of the H-band infrared spectral region, but from fitting of template spectra through the ASPCAP pipeline. As seen in Table 5, our stellar temperature and gravity are slightly lower than from the APOGEE analysis, however the differences are within 1$\sigma$. The metallicity and element abundances are in good agreement, with only Si showing a difference of up to 2$\sigma$.

This star has also been analyzed from optical spectra by \citet{Sobeck2011} and \citet{Carretta2009}. The stellar parameters from these optical analyses are in better agreement with our results, although the Fe and C abundances determined by \citet{Sobeck2011} are significantly lower than ours, and this may imply our C abundance should be lower than it's upper limit. Their C result is from a synthetic fit of the 4300 \AA, G-band, with little discussion of the errors. Throughout their analysis, \citet{Sobeck2011} report line abundances, from which we infer their mean abundances and error by adopting $\sigma/\sqrt{N}$ (unless the error was below 0.05 dex, in which case 0.05 dex was adopted). The Fe result from \citet{Carretta2009} is in good agreement with ours, however their [$\alpha$/Fe] ratios are larger by 1$\sigma$ (O) to 4$\sigma$(Si). To ensure a uniform comparison, we convert the results from each of these papers to absolute log abundances.   

\begin{table}
\centering
\begin{minipage}{80mm}
\setlength{\tabcolsep}{3pt}
	\def\arraystretch{1.0}
\caption{Standard star M15-K341: parameters and abundances}
\begin{tabular}{@{}lcccc@{}}
\hline
Stellar 	& This		& Sobeck\footnote{The abundances were retrieved using their online data table and adopting a $\sigma/\sqrt{N}$ error, except in the case where the error was below 0.05 dex, in which case 0.05 dex was adopted.}		&Carretta& M{\'e}sz{\'a}ros \\
Properties &study&et al. 2009&et al. 2009&et al. 2015\\
\hline
$\mathrm{T}_{\mathrm{eff}}$ (K)& 4282 $\pm$ 150	& 4375 $\pm$ 100	& 4324 $\pm$ 50 	& 4494 $\pm$ 100\\
log$(g)$ 					& 0.52 $\pm$ 0.20	& 0.30 $\pm$ 0.20	& 0.69 $\pm$ 0.20	& 0.82 $\pm$ 0.30 \\
Fe  &5.21 $\pm$ 0.10	&4.99 $\pm$ 0.05	& 5.32 $\pm$ 0.03 	& 5.20 $\pm$ 0.12\\
C   &$<$ 6.11	        &5.51 $\pm$ N/A	& ...				    & 5.72 $\pm$ 0.22 \\
N   & $>$ 6.27      	& ...				& ... 				& 6.13 $\pm$ 0.32\\
O   &6.96 $\pm$ 0.13	&7.02 $\pm$ 0.05	& 7.08 $\pm$ 0.11 	& 7.07 $\pm$ 0.13\\
Mg  &5.53 $\pm$ 0.19	&5.59 $\pm$ 0.10	& 5.69 $\pm$ 0.08 	& 5.63 $\pm$ 0.09\\
Al  &4.01 $\pm$ 0.21	&4.05 $\pm$ 0.13	& ... 				& 4.02 $\pm$ 0.17\\
Si  &5.45 $\pm$ 0.09	&5.30 $\pm$ 0.10	& 5.78 $\pm$ 0.08	& 5.66 $\pm$ 0.13\\
Ti  & $<$ 2.98			&2.71 $\pm$ 0.05	& ... 				& 2.68 $\pm$ 0.28\\

\hline \\[-1.0cm]
\label{table:M15}
\end{tabular}
\end{minipage}
\end{table}

\subsection{Abundance uncertainties}
\label{sec:abErr}

The sensitivities of the elemental abundances (including elements other than Fe, C, N, and O) are shown in Table \ref{table:abSens}. While M15-K341 is the standard star for this analysis, we computed these abundance sensitivities for M22-MA4.1 because this star has a temperature and metallicity that is more representative of our sample. The surface gravity of M22-MA4.1 is slightly higher than the mean of our sample, however this has a smaller impact on the precision in the elemental abundances.

The abundances in M22-MA4.1 were derived using our best model atmosphere parameters, and again with models that vary by $\pm$ 1$\sigma$ in temperature, gravity, and metallicity. The mean errors in each of the element abundances have been determined per parameter, and adopted for the other stars in this paper, as reported in Table \ref{table:abSens}. The errors in metallicity show negligible effects on the other element abundances ($<$0.2 dex), and are thus not reported here. The total errors for each species are calculated per star by combining these in quadrature with the measurement errors in Table \ref{table:abSens}.

\begin{table}
	\centering
	\caption{Abundance Uncertainties for M22-MA4.1}
	\begin{minipage}{80mm}
	\setlength{\tabcolsep}{3pt}
	\def\arraystretch{1.10}

			\begin{tabular}{@{}lcccr@{}}
			\hline
				Species			& $\Delta T_{\mathrm{eff}}$ & $\Delta$ log g & $\Delta v_t$ & Total Error\footnote{Both errors added in quadrature. Model atmosphere input metallicity sensitivities were found to be negligible and were not included here.}   \\
								& (+150K) 			    & (+0.2 dex) 	& (+0.5 km/s)	& \\
								
				\hline
				{[}Fe/H]			&-0.08		& 0.01		& 0.00	& 0.08        	\\
				{[}C/Fe]			& 0.21		& 0.00		& -0.02	& 0.21		\\
				{[}N/Fe]			& 0.04		&-0.05		& -0.01	& 0.06		\\
				{[}O/Fe]			& 0.19		&-0.02		& 0.00	& 0.19		\\
				{[}Mg/Fe]			&-0.03		& 0.02		& -0.05	& 0.06		\\
				{[}Al/Fe]			& 0.01		&-0.02		& 0.01	& 0.02		\\
				{[}Si/Fe]			& 0.01		& 0.00		& -0.03	& 0.03		\\
				{[}Ca/Fe]			&-0.04		&-0.07		& 0.00	& 0.08		\\
				{[}Ti/Fe]			& 0.12		& 0.01		& 0.00	& 0.12		\\
				{[}Mn/Fe]\footnote{Since Mn was derived as an upper limit for this star, the Mn sensitivity was computed for MA11.}			& 0.03		&-0.06		& 0.00	& 0.07		\\

				{[}Ni/Fe]			&-0.07		& 0.02		& 0.00	& 0.07		\\
				\hline
			\end{tabular}
			\end{minipage}
			\label{table:abSens}
\end{table}	


\section{Abundance results}
\label{sec:abund}

\begin{table*}
	\centering
	\caption{Target abundances}
	\begin{minipage}{180mm}
	\setlength{\tabcolsep}{5pt}
	\def\arraystretch{1.00}
	
			\begin{tabular}{@{}lcccccccccccc@{}}
			\hline

				Species			&M22-MA4		& N\footnote{Number of lines or features (i.e. band heads or molecular features) used to determine abundance.}		&M22-MA4.1		&N	&MA8			&N	&MA11			&N	&MA14			&N	&M15-K341	&N	\\
				\hline
				{[}{Fe}/H]			& -2.01 $\pm$ 0.10	& 23& -1.90 $\pm$ 0.10	&35	& -1.93 $\pm$ 0.12	&17	& -1.51 $\pm$ 0.09	&49	& -2.06 $\pm$ 0.09	&38	& -2.29 $\pm$ 0.11	&17	\\
				{[}C/Fe]			&  $<$ 0.10	        & 3	&  $<$ 0.22         &6	&  $<$ 0.35         &7	& $<$ -0.35         &6	& $<$ -0.38         &6	& $<$ -0.03         &4	\\
				{[}N/Fe]			&  $>$ 0.85         & 4	&  $>$ 0.54     	&15	&  $>$ 0.64     	&9	& $>$ 0.57      	&14	& $>$ 1.03      	&14	& $>$ 0.73      	&8	\\
				{[}O/Fe]			&  0.68 $\pm$ 0.22	&11	&  0.49 $\pm$ 0.20	&38	&  0.83 $\pm$ 0.20	&36	&  0.27 $\pm$ 0.20	&30	&  0.64 $\pm$ 0.20	&40	&  0.56 $\pm$ 0.20	&39	\\
				{[}Mg/Fe]			&  0.32 $\pm$ 0.15	&4	&  0.15 $\pm$ 0.12	&7	&  0.16 $\pm$ 0.16	&3	&  0.42 $\pm$ 0.10	&10	&  0.28 $\pm$ 0.11	&6	&  0.22 $\pm$ 0.19	&3	\\
				{[}Al/Fe]			& -0.28 $\pm$ 0.23	&2	& -0.26 $\pm$ 0.16	&3	& -0.18 $\pm$ 0.18	&2	&  0.23 $\pm$ 0.15	&3	& -0.35 $\pm$ 0.18	&2	& -0.15 $\pm$ 0.21	&2	\\
				{[}Si/Fe]			&  0.21 $\pm$ 0.09	&12	&  0.18 $\pm$ 0.08	&12	&  0.13 $\pm$ 0.09	&10	&  0.26 $\pm$ 0.07	&13	&  0.30 $\pm$ 0.09	&11	&  0.23 $\pm$ 0.09	&11	\\
				{[}S/Fe]			&  0.64 $\pm$ 0.33	&3	&  0.47 $\pm$ 0.35	&2	&  0.60 $\pm$ 0.31	&2	&  0.41 $\pm$ 0.32	&2	&  -				&-	& -				&-	\\
				{[}Ca/Fe]			&  0.28 $\pm$ 0.28	&3	&  0.05 $\pm$ 0.33	&3	&  0.29 $\pm$ 0.28	&3	&  0.28 $\pm$ 0.17	&4	&  0.27 $\pm$ 0.32	&3	&  - 				&-	\\
				{[}Ti/Fe]			&  0.65 $\pm$ 0.33	&2	&  0.30 $\pm$ 0.28	&2	&  0.26 $\pm$ 0.30	&2	&  $<$ 0.22 		&-	&  $<$ 0.27		&-	&  - 				&-	\\
				{[}Mn/Fe]			&  $<$ 0.24 		& -	& $<$ -0.05 		&-	& - 				&-	& 0.01 $\pm$ 0.32	&3	&  $<$ -0.06		&-	& $<$ -0.06 		&-	\\
				{[}Ni/Fe]			&  0.33 $\pm$ 0.21	&5	& -0.06 $\pm$ 0.19	&3	& 0.36 $\pm$ 0.25	&2	& 0.03 $\pm$ 0.16	&6	& -0.03 $\pm$ 0.25	&5	&  0.02 $\pm$ 0.34	&3	\\
{[}$\alpha$/Fe]\footnote{Taken as a weighted average between Mg, Ca, and Ti.}		&  0.36 $\pm$ 0.12	& - & 0.16 $\pm$ 0.10	&-	& 0.20 $\pm$ 0.12	&-	&  0.38 $\pm$ 0.09	&-	&  0.28 $\pm$ 0.11	&-	&  0.22 $\pm$ 0.34\footnote{In this case $\alpha$ was adopted as the weighted average between Mg and Ca only.}	&-	\\
				{[}(C+N+O)/Fe]\footnote{Error derived by adding C, N, and O abundance errors in quadrature. C is adopted here as the upper limit and its error taken as its \textit{upper} error (see Section \ref{sec:abAnal}). For simplicity we adopt this as the A(C+N+O) error as well.}	&0.58 $\pm$0.61	&-	&0.42$\pm$0.58	&-	&0.73$\pm$0.40	&-	&0.19$\pm$0.42	&-	&0.55$\pm$0.46	&-	&0.46$\pm$0.52	&-\\
				A(C+N+O)		&7.49$\pm$0.61	&-	&7.44$\pm$0.58	&-	&7.71$\pm$0.40	&-	&7.60$\pm$0.42	&-	&7.41$\pm$0.46	&-	&7.09$\pm$0.52	&- \\
				\hline
			\end{tabular}
			\end{minipage}
			\label{table:abund}
\end{table*}	

Elemental abundances have been determined from H-band infrared spectra for three stars in the Galactic bulge and two stars in M22, as a scientific demonstration of the new RAVEN instrument and MOAO methodologies.   Through the analysis of a standard star in M15, we have demonstrated that the quality of our spectra and analysis methods do reproduce previous published results.  Here we discuss the abundance results for our main science targets, also summarized in  Table \ref{table:abund}.  The results for M22 are compared with those from the optical analyses by \citet{Marino2011,Kacharov2015,Roederer2011,Mucciarelli2015}, and infrared analysis by \citet{Alves-Brito2012}; the two studies by \citet{Marino2011,Alves-Brito2012} find a large dispersion in Fe, with two rough groups spanning roughly -2 $<$ [Fe/H] $<$ -1.4 dex. Our Galactic bulge candidates are compared with bulge survey results by \citet{Koch2016,Howes2014,Howes2015,Johnson2014,GarciaPerez2013,Ruchti2011,Casey2015}.

\subsection{Iron}

{\it M22:} The mean Fe abundances for our two M22 targets is [Fe/H] = $-1.95 \pm0.05$.  This is $\sim0.2$ dex lower than the cluster average determined from the optical analyses of larger numbers of stars by \citet{Marino2011} and \citet{Mucciarelli2015}.  M22 has been found to have a spread in Fe \citep{Marino2011,Alves-Brito2012}, though the cause of this spread is debated; \citet{Mucciarelli2015} reproduce the spread only when deriving abundances from spectroscopic gravities, and show that the use of photometric gravities erases the signature. \citet{Mucciarelli2015} also argue that photometric gravities are less sensitive to systematic effects, and suggest that the spread in Fe in this cluster is not astrophysical, but a systematic effect. In this paper, we adopt photometric gravities, which would be consistent with our finding of no spread in the Fe abundances.

{\it Galactic Centre candidates:} Based on the classifications by \citet{Beers2005}, our sample includes two metal-poor and one very metal-poor star.  Before considering whether these targets are \textit{bonafide} bulge members (see orbit calculations in Section \ref{sec:orbits}), we note that they fall on the metal-poor tail of the bulge MDF.   Furthermore, these stars have metallicities between those studied by \citet{Howes2015} and those by \citet{Chiappini2011}, where each study suggests that the stars at these metallicities can provide links to the First Stars.  Our bulge candidates also fall within the metallicity range of two other works \citep{GarciaPerez2013,Koch2016}, contributing to the total sample of metal-poor stars studied in/towards the bulge. We note the discrepancy from the original EMBLA Fe abundances (determined from low resolution optical spectroscopy), which may be due to the different resolution and wavelengths between the two works; however we are confident in our abundances based on the results of our standard star comparison.

\subsection{Carbon and nitrogen}

\label{sec:carbon}
The C and N abundances were determined from the best synthetic fits to the features as described in Section \ref{sec:abAnal} and are derived as upper and lower limits, respectively.

\begin{figure}
   \centering
    \includegraphics[clip=true,trim = 0 0 0 0,width=0.48\textwidth]{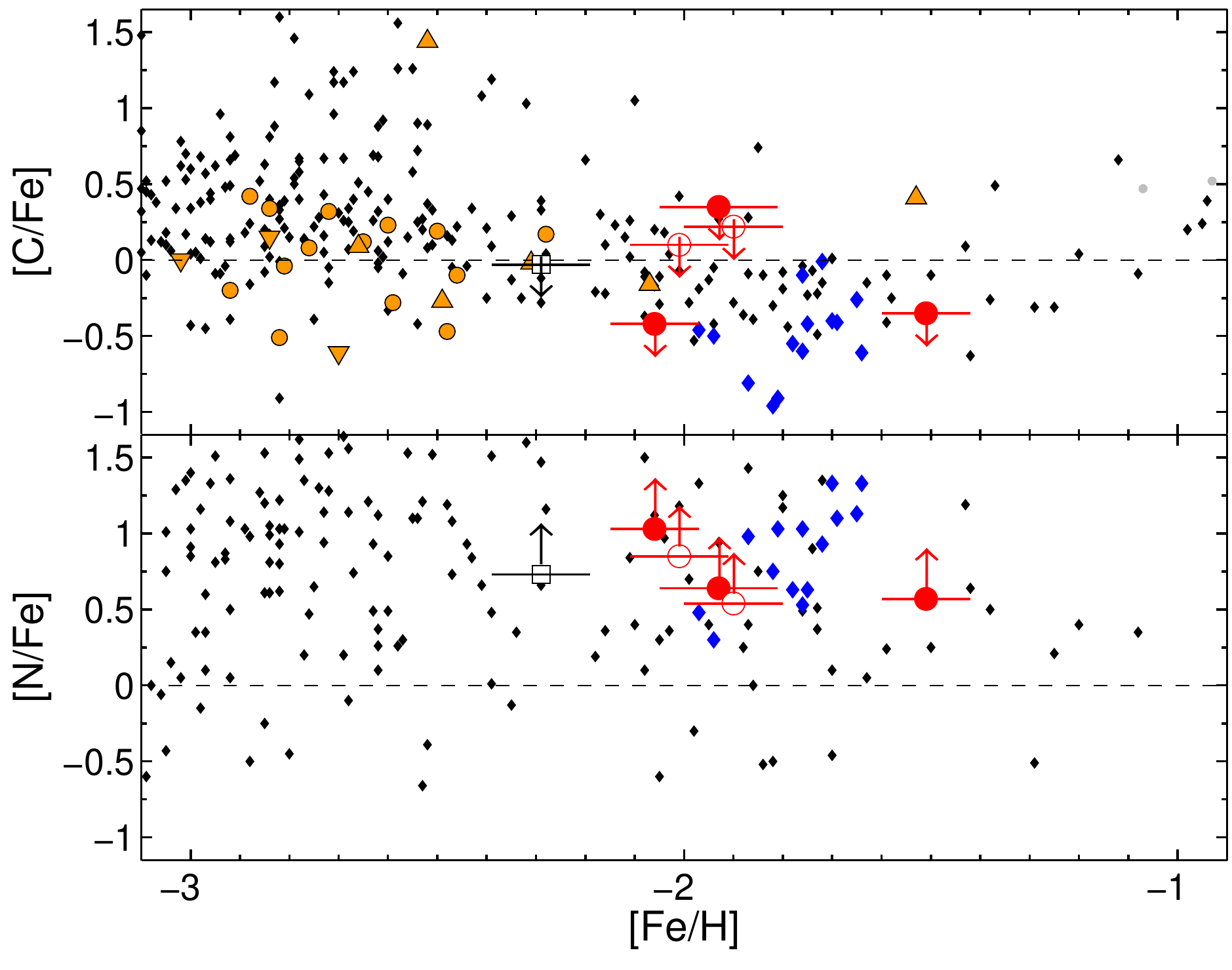}
   \caption{C and N abundances of our target stars as a function of Fe compared with the Galactic sample: thick disk taken stars from \protect\citet{Reddy2006} are shown as grey points while halo stars are shown in black (taken from \protect\citealt{Roederer2014}). Orange circles, triangles, and inverted triangles are metal-poor bulge stars from \protect\citet{Howes2015}, \protect\citet{Koch2016} and \protect\citet{Casey2015}, respectively. Blue diamonds are the abundances of 35 M22 stars, taken from \protect\citet{Marino2011} (several of their stars report multiple abundances from different observations of the same star - for these cases we adopt a straight average). The hollow and solid red points represent our M22 and Galactic Centre stars, respectively. Shown also is our standard star in M15, plotted as a hollow black square.}
   \label{fig:alphaCN}
\end{figure}

{\it M22:} Both stars in M22 show elevated N abundances coupled with slightly enhanced C relative to scaled-solar abundances. It is evident from Figure \ref{fig:alphaCN} the derived C abundances are slightly elevated with respect to \citet{Marino2011} while the N abundances between the two works agree. These abundances are consistent with mixing of CN-cycled gas if the initial abundances were slightly elevated and not simply scaled-solar (see Table \ref{table:abund}); e.g., if the initial [C/Fe] $\sim$ [N/Fe] $\sim$ +0.3 in M22 in general, then a small N enrichment is observed in our two stars with a small C depletion.   This is discussed further in Section \ref{sec:M22}.

{\it Galactic Centre candidates:} The C abundances in our metal-poor bulge candidates show [C/Fe] $<$ +0.35, relative to scaled-solar abundances.   C has been determined in other metal-poor bulge candidates by \citep{Howes2015,Koch2016,Casey2015}. Similar to \citet{Howes2015} and \citet{Casey2015}, none of our bulge candidates are very carbon-rich, which is a common feature seen in metal-poor halo stars \citet{Placco2014}. Only one bulge candidate star has been found which satisfies the CEMP criteria, as well as one that is above the halo average (\citet{Koch2016}; see Figure \ref{fig:alphaCN}). The average C abundance of the metal-poor stars from \citet{Howes2015} and \citet{Casey2015} appears to be similar to scaled-solar ([C/Fe] $\sim$ 0), slightly lower than the mean C abundance in the halo stars at similar metallicities (see Figure \ref{fig:alphaCN}). 

We report the first N abundances in metal-poor bulge candidates. They appear to be enhanced, consistent with mixing of CN-cycled gas on the RGB.   Since one of these stars has a slightly elevated C abundance, then it is unclear if they all share the same initial C and N abundances before mixing.  This is discussed further in Section \ref{sec:disc}.

\subsection{$\alpha$-elements}
\label{sec:alpha}

The individual elements that form through $\alpha$ capture processes during hydrostatic He-core burning or subsequent $\alpha$-rich freeze out in He-rich burning layers include O, Mg, Si, S, Ca, and Ti.   The results from our infrared H-band analysis are shown in Table 7, where it can be seen that measurement errors for Mg and Si are much smaller than the other elements primarily due to the larger number of lines available. We calculate a mean $<$[$\alpha$/Fe]$>$ as a weighted average of Mg, Si, Ca, and Ti.   Our [O/Fe] measurements are also sufficiently precise, however we do not average it into our $\alpha$ index.\footnote{We note that Mg can also be affected through other processes in globular clusters, however our two stars in M22 have identical Al, implying that deep mixing has not affected Mg and so we keep Mg in our computation of $\alpha$.}

The average $\alpha$ abundances for our target stars are compared with the galactic sample in Figure \ref{fig:alphaB}. For the comparison sources, [$\alpha$/Fe] was computed as an unweighted average of Mg, Si, Ca, and Ti except in the cases of \citet{GarciaPerez2013,Howes2014,Johnson2013} where each source contained only 3 out of 4 of these elements, and the alpha abundance was computed as the average of these 3. 
Our M22 target abundances are consistent with those of \citet{Marino2011}, with a slightly lower average of $<$[$\alpha$/Fe]$>$~=~0.22 $\pm$ 0.10 compared to their $<$[$\alpha$/Fe]$>$~=~0.33 $\pm$ 0.04. For the three Galactic bulge stars, we report $<$[$\alpha$/Fe]$>$~=~0.25 $\pm$ 0.11, which is in excellent agreement with the mean abundances $<$[$\alpha$/Fe]$>$~=~ 0.27 $\pm$ 0.13 from \citet{Howes2014,Howes2015}, also in agreement with \citet{GarciaPerez2013} who find $<$[$\alpha$/Fe]$>$~=~ 0.26 (no reported error).
Inspection of Figure \ref{fig:alphaB} reveals this is in slight contrast to the higher $\alpha$ abundances reported by \citet{Casey2015,Koch2016}, and the metal-poor stars taken from \citet{Johnson2014}.

\subsubsection{Oxygen:} 
Most O abundances derived in the literature have been inferred from optical spectroscopy, usually from either the forbidden line ($\sim$6300 \AA) or from the triplet region between 7770-7777 \AA. These lines are notoriously difficult to analyse at low [Fe/H] due to 3D effects, departures from LTE, small line strengths and/or blends (e.g. \citealt{Asplund2005,Amarsi2016}). Relatively few O abundance determinations are thus available for our comparison sample (Fig. 6). Since the H-band is host to a plethora of OH lines, it is an excellent wavelength region to determine stellar O abundances.  We have derived O abundances from a selection of 58 OH lines of varying line strengths.

From two stars in M22, we find [O/Fe]~= 0.68 $\pm$ 0.22 and 0.49 $\pm$ 0.20, which is significantly higher than \citet{Marino2011} and \citet{Alves-Brito2012}, where $<$[O/Fe]$>$~=~0.34 $\pm$ 0.03 (from the single forbidden line at 6300 \AA) and 0.40 (no reported error), respectively. 
\citet{Alves-Brito2012} also report a large spread in O ($\Delta$log(O/Fe)=0.74); such a spread in globular clusters has been explained as proton capture reactions which impact O and Mg \citep{Marino2011}. A strong Na-O anticorrelation has been determined in this cluster \citet{Marino2011}; our high O abundances relative to the sample of M22 stars (see Figure \ref{fig:alphaA}) suggest that our two M22 stars represent primordial stars within the cluster.

Oxygen abundances for bulge stars have been rarely determined in this metallicity range (as seen in Figure \ref{fig:alphaA}). For our three bulge candidates we find $<$[O/Fe]$>$~= 0.58 $\pm$0.20. This is in excellent agreement with both \cite{GarciaPerez2013} ($<$[O/Fe]$>$~=0.54 $\pm$ 0.07 from 5 stars) and \citet{Johnson2014} ($<$[O/Fe]$>$~=~0.61 $\pm$ 0.07, from 3 metal-poor stars with [Fe/H] $<$ -1 dex). Similar to previous results, this averaged value is higher than metal-poor stars in the halo, but in good agreement with the thick disk stars \cite[e.g.,][]{Roederer2014,Johnson2014} (we caution however that the thick disk stars should generally agree with the halo stars, and the offset between the two in Figure \ref{fig:alphaA} may merely be a systematic effect).   

\subsubsection{Magnesium:} 
Magnesium is measured from several prominent lines in the IR, the most notable of which are the 3 at 15025-15050\AA, and the 3 at 15740-15770\AA, and several weaker lines spread throughout the H-band. We find $<$[Mg/Fe]$>$~=~0.24 $\pm$0.14 for our two M22 targets.  Figure \ref{fig:alphaA} shows these abundances relative to Galactic globular clusters and results from the optical analysis of stars in M22 by \citet{Marino2011}.  Our results are slightly lower than the average from \citet{Marino2011}, where $<$[Mg/Fe]$>$~=~0.39 $\pm$ 0.02 dex. Our three Galactic Centre candidates have $<$[Mg/Fe]$>$~=~0.29 $\pm$ 0.13 (see Figure \ref{fig:alphaA}).  Like \citet{GarciaPerez2013} and \citet{Howes2015}, these Mg abundances are lower than the optical results for stars of similar metallicity in the halo.  This is also in contrast to results by \citet{Koch2016} and \citet{Johnson2014}, who found higher [Mg/Fe] from optical analyses of stars in the bulge and thick disk.  \citet{Howes2014} report a large spread of Mg $\Delta$[Mg/Fe] = 0.69, while \citet{Casey2015} find little to no Mg variation in their sample.  Our data would support a dispersion in the bulge star Mg abundances.

\subsubsection{Silicon:}
\begin{figure}
   \centering
    \includegraphics[clip=true,trim = 0 0 0 0,width=0.48\textwidth]{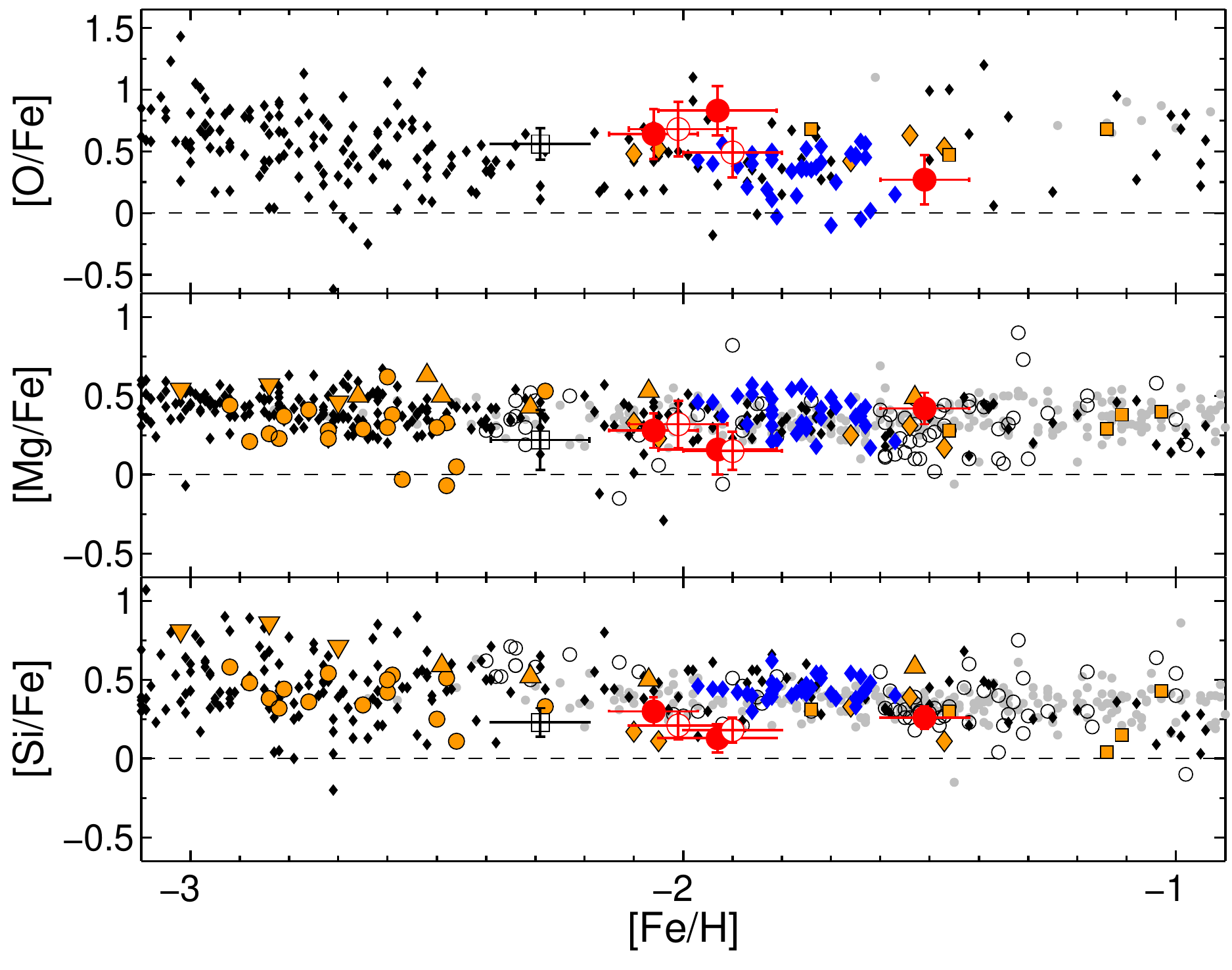}
   \caption{The light elements O, Mg, and Si plotted as a function of Fe compared with the Galactic sample. The grey points are thick disk stars from \protect\citet{Reddy2006} and \protect\citet{Ruchti2011} while the black points (solid) represent halo stars (taken from \protect\citet{Roederer2014} and \protect\citet{Reddy2006}). Also shown are Galactic GCs as hollow black circles, assembled by \protect\citet{Pritzl2005} and metal-poor bulge stars in orange (diamonds \protect\citet{GarciaPerez2013}, squares \protect\citet{Johnson2014}, triangles \protect\citet{Koch2016}, inverted triangles \protect\citealt{Casey2015} and circles \protect \citealt{Howes2014,Howes2015}). Blue diamonds are the abundances of 35 M22 stars, taken from \protect\citet{Marino2011} (several of their stars report multiple abundances from different observations of the same star - for these cases we adopt a straight average). The M22 and Galactic Centre targets from this work red open and filled circles, respectively while the standard star in M15 is represented by a hollow black square.}
   \label{fig:alphaA}
\end{figure}

We derive Si abundances from numerous lines of varying strength ($\ge$10 lines per star). The average $<$[Si/Fe]$>$~=~0.20 $\pm$0.09 for our two M22 stars which, similar to oxygen, is lower than \citet{Marino2011} (where $<$[Si/Fe]$>$~=~0.44 $\pm$ 0.01 dex); this is readily seen by inspection of Figure \ref{fig:alphaA}.
Our M22 Si abundances agree with other Galactic globular clusters at this metallicity (see Figure \ref{fig:alphaA}). We find $<$[Si/Fe]$>$~=~0.23 $\pm$0.09 for our three bulge candidates, in agreement with most other bulge studies (\citealt{GarciaPerez2013}, \citealt{Johnson2013}, and \citealt{Howes2015}); only \citet{Koch2016} and \citet{Casey2015} find higher Si results (see Figure \ref{fig:alphaA}).

\label{sec:alpha}

\subsubsection{Calcium:}
Calcium is measured from the 4 features located between 16150-16200 \AA.  It is clear from Table 7 that our infrared Ca abundances have large uncertainties, due to a large line-to-line scatter. We were unable to measure the Ca lines in the M15 standard star, thus it is difficult to critically evaluate the 4 individual spectral lines used in this analysis. The resulting abundances are plotted in Figure \ref{fig:alphaB} along with our comparison sample. The average Ca for our two stars in M22 is $<$[Ca/Fe]$>$~=~0.17 $\pm$0.30. This is in fair agreement with results from \citet{Marino2011} ($<$[Ca/Fe]$>$~=~0.30 $\pm$ 0.01) given our large uncertainties. The mean Ca abundance in our sample of bulge stars is $<$[Ca/Fe]$>$~=~0.28 $\pm$0.26, which is consistent with the 26 bulge stars in \cite{Howes2015} ($<$[Ca/Fe]$>$~=~0.23 $\pm$ 0.12). \citet{Johnson2014}, \citet{Koch2016}, and \citet{Casey2015} find slightly higher Ca for bulge stars ($<$[Ca/Fe]$> \sim 0.4$), though all of these analyses are within 1$\sigma$ errors. We note that \citet{Howes2015} find lower Ca in their bulge sample than halo stars at the same metallicities, though the source of this discrepancy is unclear.

\subsubsection{Titanium:}
\begin{figure}
   \centering
    \includegraphics[clip=true,trim = 0 0 0 0,width=0.48\textwidth]{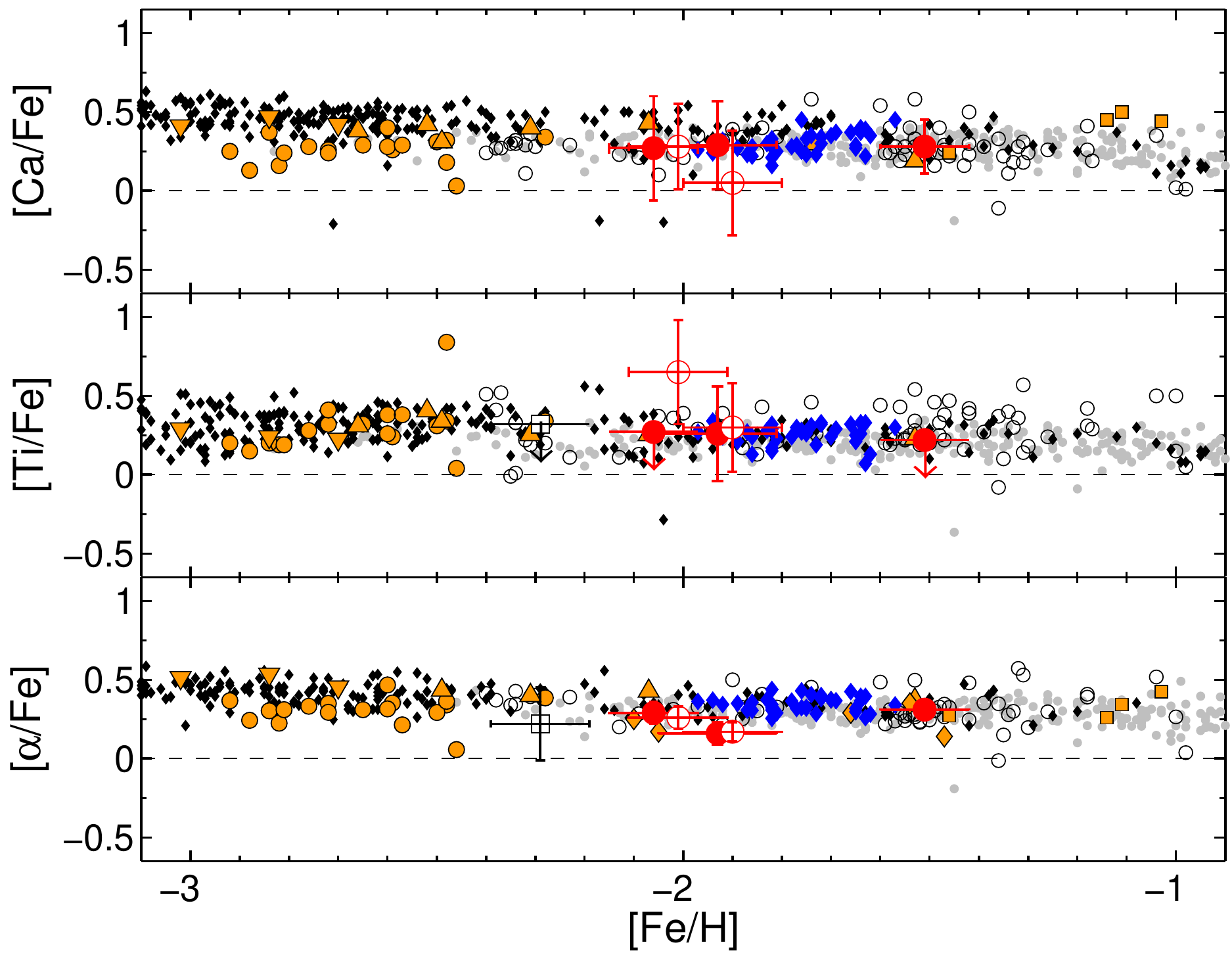}
   \caption{The light elements Ca, Ti, and $\alpha$ plotted as a function of Fe compared with the Galactic sample. Data points are labelled the same as in Figure \ref{fig:alphaA}. The Ti abundances from \protect\citet{Roederer2014} (solid black points), \protect\citet{Koch2016} (orange triangles), and \protect\citet{Ruchti2011} (light gray circles) are taken as an average between Ti I and Ti II.}
   \label{fig:alphaB}
\end{figure}

Titanium is measured from two features in our IR spectra,  15335 \AA (which is very close to a strong Fe feature) and a weak line at 15544 \AA. Inspection of Figure \ref{fig:alphaB} indicates that our Ti abundances for the two M22 stars are quite high, with $<$[Ti/Fe]$>$~=~0.48 $\pm$0.31, compared with \citet{Marino2011}, however with a large uncertainty. These results are also consistent with the Galactic sample of globular clusters. Ti is measured in only one Galactic bulge candidate (MA8), and determined with upper limits for the other two. The abundances we report in Table \ref{table:abund} are consistent with those for other bulge stars, as well as the thick disk and halo stars which overlap at this metallicity (see Figure \ref{fig:alphaB}).

\subsubsection{Sulphur:}
We examine three S I features that span the range 15400-15480 \AA.  These three features yield significantly different results, and without detections of these lines in our M15 standard star then it is difficult to critically evaluate these results.   We report the mean abundance for [S/Fe] here, however the unrecognized blends or poor atomic data could be affecting these infrared S I abundances.
We calculate $<$[S/Fe]$>$~=~0.56 $\pm$ 0.48 dex from two stars in M22; this is similar to \citet{Kacharov2015}, who find $<$[S/Fe]$>$~=~0.57 $\pm$0.21 from an average of 6 stars in the same cluster (also with comparably large errors).  

Sulphur is measured in 2 out of 3 of our Galactic bulge candidates, where $<$[S/Fe]$>$~=~0.51 $\pm$0.32.  These are the first S I abundances reported for Galactic bulge candidates, but the uncertainties are too high to discuss the nucleosynthesis or S I compared with Galactic halo stars. Detailed analyses of S I measurements have been carried out by \citet{Nissen2007,Jonsson2011} and find $<$[S/Fe]$>$~=~0.20 $\pm$ 0.07, and 0.40 $\pm$ 0.11 respectively, which is within the 1$\sigma$ errors of our S I abundances.

\subsection{Other elements}
\label{sec:others}
Three more elements are available in our H-band spectra: the odd-elements Al and Mn, and Ni.   Odd-elements typically are affected by hyperfine splitting, which is not yet available for these IR lines.   Elemental abundances for these three elements are shown in Figure 8.

\subsubsection{Aluminum:}

In the H-band, there are three prominent Al features, located between 16718 - 16765 \AA, from which we derive our Al abundances. \citet{Shetrone2015,Smith2013} chose to not include the strongest feature at 16750 \AA, because of potential HFS effects. However we found no significantly large abundance difference with the weaker Al features. In M22, we search for a Mg-Al anti-correlation in the discussion. We find $<$[Al/Fe]$>$~=~-0.27 $\pm$ 0.20, which may be anti-correlated with our slightly elevated [Mg/Fe] results. High Al was found in \citet{Marino2011} which we do not reproduce, and we attribute our low Al abundances as evidence of these stars being of the first generation in the cluster. Inspection of Figure \ref{fig:alphaC} reveals that the galactic sample at these low metallicities is slightly lower than the abundances of both our M22 and Galactic bulge targets. We report $<$[Al/Fe]$>$~=~-0.10 $\pm$ 0.29 for our three Galactic Centre candidates, where most of the spread is attributed to the higher Al abundance in MA11 (however MA11 agrees with the Al trends of other stars at its metallicity). Only \citet{Johnson2013} have measured Al in a galactic bulge star at our target metallicities, for which we find excellent agreement (see Figure \ref{fig:alphaC}).

\subsubsection{Manganese:}
Three Mn I features are examined, at 15159, 15218, and 15262 \AA, and primarily used to determine [Mn/Fe] upper limits. These upper limits for our two M22 targets, show [Mn/Fe] $\le$ 0.2. This is consistent with \citet{Roederer2011}, who derive $<$[Mn/Fe]$>$~=~-0.52 $\pm$ 0.05 dex from 6 RGB stars in M22 (we computed the error from a weighted average of their reported abundances). For the Galactic bulge targets, the upper limits and one detection result in scaled-solar abundances, with [Mn/Fe] $\le$ 0. \cite{Casey2015} find underabundant [Mn/Fe] for their bulge star sample; we do not reproduce this feature, similar to \cite{Howes2016}.

\subsubsection{Nickel:}
There are several Ni features in the H-band. We determine Ni abundances from $\le$6 lines spanning across the spectrum (from 15555 - 16996 \AA). In M22, our two stars yield very different [Ni/Fe] results, but with large uncertainties; $<$[Ni/Fe]$>$~=~0.14 $\pm$0.28. This is consistent with \citet{Marino2009} ($<$[Ni/Fe]$>$~=~-0.07 $\pm$0.04). In our three Galactic bulge candidates, two stars show scaled-solar nickel, with [Ni/Fe] $\sim$ 0, while the third appears Ni rich with [Ni/Fe] = 0.36 $\pm$0.25; this star clearly stands out from the Galactic comparison stars in Figure 8.

\begin{figure}
   \centering
    \includegraphics[clip=true,trim = 0 0 0 0,width=0.48\textwidth]{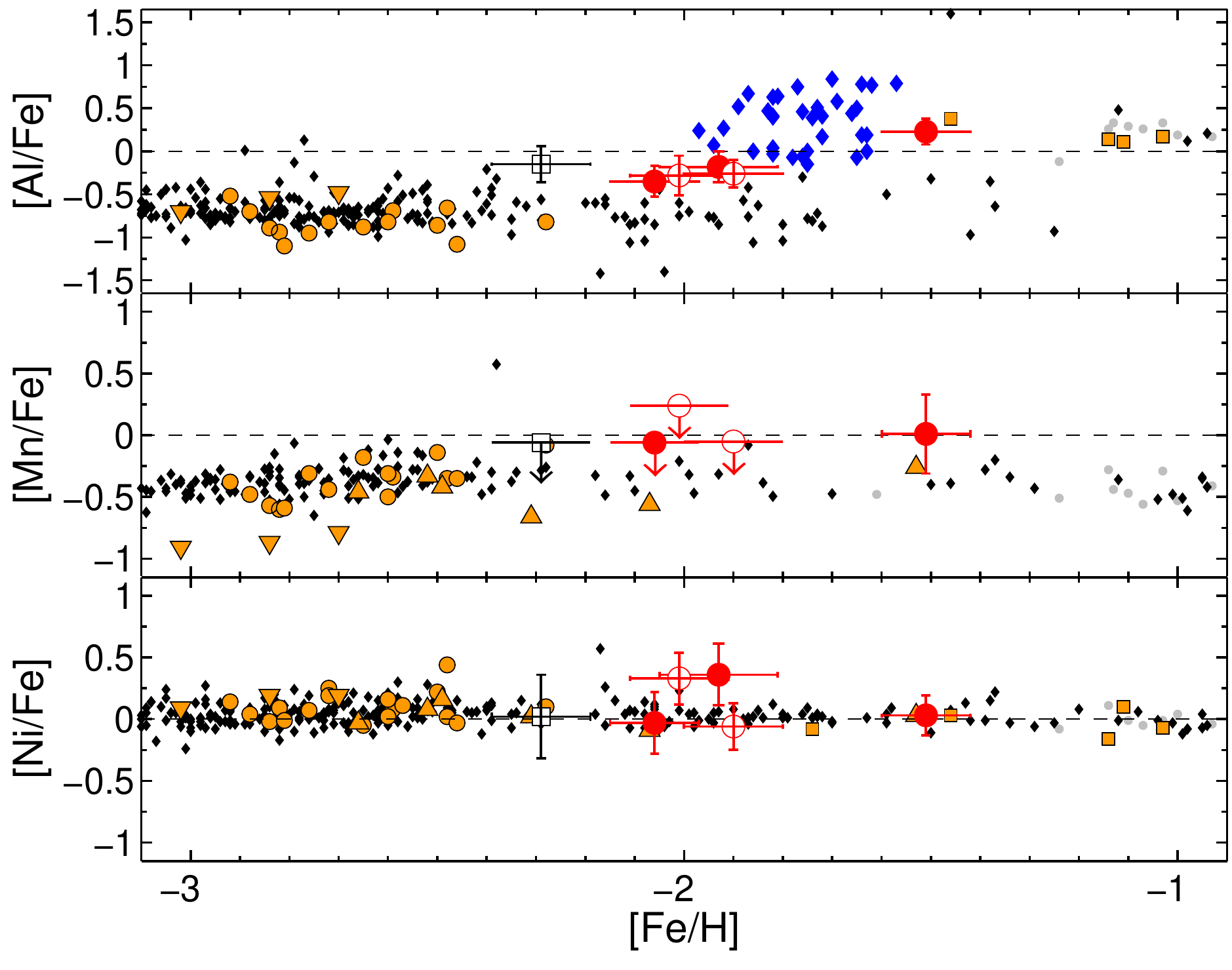}
   \caption{The light elements Al, Mn, and Ni plotted as a function of Fe compared with the Galactic sample. Data points are labelled the same as in Figure \ref{fig:alphaA}.}
   \label{fig:alphaC}
\end{figure}

\section{Stellar orbits}

The Galactic bulge candidate stars may be transient in nature (i.e. foreground disk/halo, or halo stars with orbits that pass through the bulge); therefore they need to be identified as \textit{bonafide} bulge members if they are to be used to study the evolution of the bulge stellar population. To affirm their bulge membership, we determine their orbital parameters by calculating orbits from stellar distances, proper motions, positions on the sky, and radial velocities.

\label{sec:orbits}

\subsection{Distances}
\label{sec:distances}
The distances to our target stars are determined from the absolute magnitudes of each star, based on their stellar parameters ($T_\mathrm{eff}$ and log~g), bolometric magnitudes, and bolometric corrections ($BC_V$), following \citet{Buzzoni2010}. The bolometric magnitudes were calculated assuming a stellar mass of 0.8 $M_{\odot}$.  These distances are corrected for reddening using the maps of \citet{Schlegel1998}; the computed heliocentric distances are listed in Table \ref{table:orbitalParams}. Upper and lower errors on the calculated distances are determined by varying the uncertainties in both the stellar parameters and the reddening values; we adopt the absolute lowest and highest values as our errors. As a comparison, the distance to M22 is determined from the distance modulus of \citet{Monaco2004}, $D_{M22} = 3251^{+313}_{-286}$ pc. This is consistent with our two stars, $D_{\mathrm{MA4}} = 3479^{+1314}_{-978}$ and $D_{\mathrm{MA4.1}} = 3432^{+1328}_{-987}$ pc. The final derived distances and errors for all target stars are summarized in Table \ref{table:orbitalParams}.

\subsection{Proper Motions and Stellar Kinematics}

Assuming a Milky Way potential, the stellar kinematics of the bulge candidates can be calculated from their distances, radial velocities, proper motions, and positions on the sky. Using the distances and radial velocities derived in this work, along with the positions of the stars, we examine the proper motions from both UCAC4 \citep{Zacharias2012} and SPM4 \citep{Girard2011}; and orbits are represented by their apocentre and pericentre distances. For this calculation, heliocentric parameters are transformed to Galactocentric distances and velocities assuming the LSR circular motion of 220 km/s, the Sun location of 8.5 kpc from the Galactic centre, and ($U_{\rm \odot}$,$V_{\rm \odot}$,$W_{\rm \odot}$)\footnote{(U,V,W) is a right handed system, with U pointing towards Galacticcentre, V to the direction of Galactic rotation, and W to the North Galactic Pole.}=(10.1,4.0,6.7) km/s \citep{Hogg2005} for peculiar motion of the Sun. Uncertainties in the LSR and solar motion measurements are negligible compared to the uncertainties in the distance and velocities of our stars. It should be noted that the error bars on the proper motions can be quite large, and can differ significantly from one catalogue to the next (see Table \ref{table:orbitalParams}). These discrepancies translate to large orbit uncertainties, and highlight the need for more reliable proper motion measurements (i.e. GAIA), and we caution the reader about the orbits calculated with these proper motions.  The apocentre and pericentre of each star are computed using both proper motion catalogues, using two different methods; each method is described below.

\begin{table*}
 \setlength{\tabcolsep}{7pt}
	\def\arraystretch{1.20}
\caption{bulge Candidate Observed Parameters}
\begin{tabular}{@{}lcccccc@{}}
\hline
&       &                                               &\multicolumn{2}{c}{SPM4 \citep{Girard2011}}&\multicolumn{2}{c}{UCAC4 \citep{Zacharias2012}}\\
\cmidrule(r){4-5}\cmidrule(r){6-7}
Star 	& $r_{\odot}$	        &V$_{hel}$              & $\mu_{\alpha} \cos{\delta}$ 	& $\mu_{\delta}$	& $\mu_{\alpha} \cos{\delta}$ 	& $\mu_{\delta}$\\
		&	(kpc)	            &km/s	                &	(mas/yr)	    &	(mas/yr)		    &(mas/yr)	&	(mas/yr)\\
\hline
MA8		& $4.5^{+1.8}_{-1.3}$   &-215.59 $\pm$ 0.98        & -3.93	$\pm$ 3.96	    & -10.86 $\pm$ 3.98     & -13.9 $\pm$ 2.5 	&-4.0	$\pm$ 3.4 \\
MA11 	& $11.0^{+4.3}_{-3.2}$	& 113.06 $\pm$ 0.45        & -5.22	$\pm$ 2.62	    & -7.86 $\pm$ -2.64     &-3.4	$\pm$ 2.1   &-13.6	$\pm$ 2.1 \\
MA14	& $4.5^{+1.9}_{-1.4}$	&-136.66 $\pm$ 0.58        & -1.23	$\pm$ 4.08	    & -5.67 $\pm$ 4.11      & 0.1   $\pm$ 4.5   &-2.9   $\pm$ 4.6\\
\hline \\[-0.5cm]
\label{table:orbitalParams}
\end{tabular}
\end{table*}

\subsubsection{Method 1: APOSTLE Potential Models}
\label{sec:aziOrbits}

For the potential of the Milky Way, we adopt {\it spherically averaged} potentials of Milky Way-size haloes from the APOSTLE\footnote{A Project Of Simulating The Local Environment} project \citep{Fattahi2016, Sawala2016}, a suite of high resolution hydrodynamical zoom-in simulations of twelve volumes resembling the Local Group. The simulations were performed using the EAGLE galaxy formation code \citep{Schaye2015,Crain2015}. The interested reader is referred to \citet{Fattahi2016} and \citet{Sawala2016} for the details of the runs and the selection of Local Group candidates. Simulations have been set up at three different resolutions. In this paper, we use six Milky Way-size haloes from the three highest resolution simulations \citep[AP-1, AP-4, and AP-11 in ][]{Fattahi2016} of APOSTLE. No other haloes were simulated at this resolution and so only these six were used in this study.

Given the potentials and Galactocentric distances and velocities of bulge candidate stars, we compute the apocentre and pericentre of their orbits. Uncertainties in apocentre and pericentre were obtained using 1000 Monte Carlo experiments, assuming Gaussian errors in observed heliocentric distances and velocities. For distance measurements with asymmetric errors, we assume Gaussians of different widths on either side of the mean value. Table \ref{table:orbitalParams2} summarizes the results, where 6 rows for each star correspond to 6 Milky Way analog hosts, and lower/upper errors represent $\pm 1\sigma$ errors.

\subsubsection{Method 2: Galpy}

bulge candidate orbits were also computed using Galpy, a python and c-based code made freely available\footnote{https://github.com/jobovy/galpy} by Jo Bovy. We adopt the standard Milky Way potential described in \citet{Bovy2015}, consisting of a power-law profile bulge, Miyamoto-Nagai disk, and Navarro-Frenk-White halo \citep{Miyamoto1975,NFW1996}. Median apocentres and pericentres of each star are computed by integrating their orbits using the aforementioned 3D potential model and orbital parameters in Table \ref{table:orbitalParams}; these results assume the same galactic circular velocity, radius, and (U,V,W) as in our APOSTLE calculations (see Section \ref{sec:aziOrbits}). The results are presented in Table \ref{table:orbitalParams3}. Finally, 1000 orbits were computed (following \citealt{Howes2015}), using a Monte Carlo simulation along with the uncertainties from Table \ref{table:orbitalParams} to determine 1-sigma errors. These orbital errors are in Table \ref{table:orbitalParams}.

\subsection{Orbits}

\label{sec:orbitConclusion}

A recent study using 2MASS data \citep{Robin2012} find a two component model best fits the shape of the Galactic bulge: one component is physically smaller and more metal rich, with a major-axis scale length of 1.46$\pm$ 0.25 kpc, while the second component is less massive, more metal-poor, and physically larger, with scale length of 4.44 $\pm$ 0.25 kpc. While the first component is attributed to the main boxy bar of the Galactic Centre, they suggest the second component may represent a flattened classical bulge, driven to this shape by the potential of the bar. If we consider our bulge candidates as potential members of the latter metal-poor component (given their metallicities) and we find their orbit apocentres are within one scale length, or 4.44 $\pm$ 0.25 kpc, then they can be safely assumed as bulge members. \citet{Robin2012} also find the bulge cutoff radius of $\sim$ 6 kpc.

\begin{itemize}
\item \textit{MA8:} In no calculation does the bulge candidate MA8 show an orbit constrained to the Galactic bulge; this suggests MA8 is a transient halo member temporarily passing through the centre of the Milky Way.
\item \textit{MA11:} Lower bound apocentres calculated using both the SPM4 proper motions and Method 2 (using Galpy) show an apocentre of 3.2 kpc - well within the Galactic bulge. The smallest lower bound apocentres for Method 1 (MW1 and MW4 potentials), yields 4.8 and 5.5 kpc (respectively) - marginally outside one scale length yet still within the bulge cutoff radius. Neither method yields realistic bulge orbits when considering the UCAC4 proper motions, emphasizing the need for high quality proper motions to alleviate this discrepancy (e.g. soon to be available from GAIA \citealt{GAIA}). We conclude the kinematics of MA11 suggest it may be a potential bulge member, however the majority of the orbit calculations do not support this.
\item \textit{MA14:} Using SPM4 proper motions, a lower bound apocentre is found within one scale length using Method 1 (MW1 potential) and Method 2 (found to be 4.5 and 3.6 kpc, respectively). Lower bound apocentres for five out of the six MW-like potentials are found to be $<6$ kpc - within the bulge cutoff radius. Method 2 calculated with the UCAC4 proper motions suggests MA14 is a bulge member, with a lower bound apocentre of 4 kpc; however Method 1 produces a larger lower bound apocentre of 5.5 kpc (but still within the bulge cutoff radius). Therefore, the kinematics of MA14 show roughly half of the calculated orbits have lower bound apocentres consistent with bulge membership, and we conclude this star is a potential bulge member.

\end{itemize}

\begin{table*}
 \begin{minipage}{160mm}
 \setlength{\tabcolsep}{3pt}
	\def\arraystretch{1.30}
\caption{bulge Candidate Orbits using APOSTLE MW-like Potentials}
\begin{tabular}{@{}lccccc@{}}
\hline
 && \multicolumn{2}{c}{SPM4 \citep{Girard2011} proper motions}& \multicolumn{2}{c}{UCAC4 \citep{Zacharias2012} proper motions} \\
 \cmidrule(r){3-4}\cmidrule(r){5-6}
	&		&	Median						&	Median						&	Median						&	Median						\\
	&	Potential\footnote{Each potential is a MW-like, taken from the APOSTLE project. See Section \ref{sec:aziOrbits} for a more detailed description.}	&	apocentre (kpc)						&	pericentre (kpc)					&	apocentre	(kpc)					&	pericentre	(kpc)					\\
\hline
MA8	&	MW1	&	15.3	$_{-	5.4	}^{+	12.1	}$	&	1.8	$_{-	0.9	}^{+	0.8	}$	&	18.5	$_{-	5.6	}^{+	17.3	}$	&	3.1	$_{+	0.9	}^{+	0.6	}$	\\
	&	MW2	&	41.6	$_{-	20.6	}^{+	99.6	}$	&	1.9	$_{-	0.9	}^{+	0.9	}$	&	62.2	$_{-	30.2	}^{+	191.9	}$	&	3.2	$_{+	0.9	}^{+	0.7	}$	\\
	&	MW3	&	20.2	$_{-	7.5	}^{+	23.2	}$	&	1.8	$_{-	0.8	}^{+	0.9	}$	&	25.9	$_{-	9.0	}^{+	38.0	}$	&	3.2	$_{+	1.0	}^{+	0.6	}$	\\
	&	MW4	&	23.1	$_{-	10.3	}^{+	38.8	}$	&	1.8	$_{-	0.9	}^{+	0.9	}$	&	32.2	$_{-	13.6	}^{+	66.7	}$	&	3.1	$_{+	0.9	}^{+	0.7	}$	\\
	&	MW5	&	23.6	$_{-	9.9	}^{+	40.5	}$	&	1.8	$_{-	0.8	}^{+	0.9	}$	&	32.6	$_{-	13.5	}^{+	75.6	}$	&	3.2	$_{+	1.0	}^{+	0.6	}$	\\
	&	MW6	&	28.1	$_{-	11.9	}^{+	50.4	}$	&	1.9	$_{-	0.9	}^{+	0.9	}$	&	39.1	$_{-	16.4	}^{+	100.9	}$	&	3.2	$_{+	0.9	}^{+	0.7	}$	\\
MA11	&	MW1	&	18.8	$_{-	14.0	}^{+	>300	}$	&	2.9	$_{-	1.8	}^{+	3.9	}$	&	246.4	$_{-	233.8	}^{+	>300	}$	&	3.1	$_{+	1.5	}^{+	3.9	}$	\\
	&	MW2	&	75.2	$_{-	66.5	}^{+	>300	}$	&	2.9	$_{-	1.7	}^{+	4.0	}$	&	>300	$_{-	>300	}^{+	>300	}$	&	3.1	$_{+	1.5	}^{+	3.9	}$	\\
	&	MW3	&	28.1	$_{-	22.0	}^{+	>300	}$	&	2.9	$_{-	1.7	}^{+	3.9	}$	&	>300	$_{-	>300	}^{+	>300	}$	&	3.1	$_{+	1.5	}^{+	3.9	}$	\\
	&	MW4	&	34.2	$_{-	28.7	}^{+	>300	}$	&	2.9	$_{-	1.8	}^{+	3.9	}$	&	>300	$_{-	>300	}^{+	>300	}$	&	3.1	$_{+	1.5	}^{+	3.9	}$	\\
	&	MW5	&	36.4	$_{-	30.1	}^{+	>300	}$	&	2.9	$_{-	1.7	}^{+	3.9	}$	&	>300	$_{-	>300	}^{+	>300	}$	&	3.1	$_{+	1.5	}^{+	3.9	}$	\\
	&	MW6	&	45.0	$_{-	37.5	}^{+	>300	}$	&	2.9	$_{-	1.7	}^{+	3.9	}$	&	>300	$_{-	>300	}^{+	>300	}$	&	3.1	$_{+	1.5	}^{+	3.9	}$	\\
MA14	&	MW1	&	7.2	$_{-	2.7	}^{+	4.6	}$	&	1.9	$_{-	1.0	}^{+	1.4	}$	&	9.7	$_{-	4.2	}^{+	8.8	}$	&	2.4	$_{+	1.2	}^{+	1.5	}$	\\
	&	MW2	&	12.3	$_{-	5.7	}^{+	16.0	}$	&	2.2	$_{-	1.1	}^{+	1.5	}$	&	21.1	$_{-	12.7	}^{+	42.2	}$	&	2.7	$_{+	1.4	}^{+	1.4	}$	\\
	&	MW3	&	8.6	$_{-	3.3	}^{+	6.8	}$	&	2.1	$_{-	1.1	}^{+	1.4	}$	&	12.4	$_{-	6.0	}^{+	13.6	}$	&	2.6	$_{+	1.3	}^{+	1.4	}$	\\
	&	MW4	&	8.4	$_{-	3.5	}^{+	8.1	}$	&	2.0	$_{-	1.1	}^{+	1.4	}$	&	12.7	$_{-	6.6	}^{+	19.8	}$	&	2.5	$_{+	1.3	}^{+	1.5	}$	\\
	&	MW5	&	9.0	$_{-	3.6	}^{+	8.1	}$	&	2.1	$_{-	1.1	}^{+	1.4	}$	&	13.5	$_{-	6.9	}^{+	19.3	}$	&	2.6	$_{+	1.3	}^{+	1.4	}$	\\
	&	MW6	&	10.3	$_{-	4.3	}^{+	10.1	}$	&	2.1	$_{-	1.0	}^{+	1.5	}$	&	16.0	$_{-	8.5	}^{+	23.6	}$	&	2.7	$_{+	1.4	}^{+	1.4	}$	\\
\hline \\[-0.5cm]
\label{table:orbitalParams2}
\end{tabular}
\end{minipage}
\end{table*}

\begin{table*}
 \begin{minipage}{160mm}
 \setlength{\tabcolsep}{5pt}
	\def\arraystretch{1.30}
\caption{bulge Candidate Orbits using Galpy}
\begin{tabular}{@{}lccccccr@{}}
\hline

	&	PM\footnote{Proper Motion sources: \citet{Girard2011} (SPM4) and \citet{Zacharias2012} (UCAC4).}	&	Median						&	Median						&	Median						&	Median						\\
	&	source	&	apocentre (kpc)						&	pericentre (kpc)						&	ecc.						&	$Z_{max}$						\\
	\hline
MA8	&	SPM4	&	9.22	$_{-	3.02	}^{+	5.51	}$	&	1.08	$_{-	0.65	}^{+	0.95	}$	&	0.79	$_{-	0.11	}^{+	0.12	}$	&	2.91	$_{-	2.08	}^{+	6.16	}$	\\
	&	UCAC4	&	9.93	$_{-	2.62	}^{+	5.72	}$	&	2.06	$_{-	1.03	}^{+	0.84	}$	&	0.66	$_{-	0.07	}^{+	0.17	}$	&	7.93	$_{-	3.55	}^{+	5.39	}$	\\
MA11	&	SPM4	&	9.22	$_{-	6.02	}^{+	492.94	}$	&	2.09	$_{-	1.53	}^{+	3.98	}$	&	0.77	$_{-	0.28	}^{+	0.21	}$	&	4.89	$_{-	3.21	}^{+	80.46	}$	\\
	&	UCAC4	&	585.98	$_{-	579.54	}^{+	2083.26	}$	&	3.34	$_{-	2.33	}^{+	3.76	}$	&	0.99	$_{-	0.28	}^{+	0.01	}$	&	274.38	$_{-	271.03	}^{+	1028.85	}$	\\
MA14	&	SPM4	&	5.50	$_{-	1.89	}^{+	2.46	}$	&	1.30	$_{-	0.86	}^{+	1.31	}$	&	0.60	$_{-	0.14	}^{+	0.22	}$	&	0.89	$_{-	0.39	}^{+	1.30	}$	\\
	&	UCAC4	&	6.51	$_{-	2.44	}^{+	4.00	}$	&	1.89	$_{-	1.29	}^{+	1.43	}$	&	0.57	$_{-	0.12	}^{+	0.21	}$	&	1.11	$_{-	0.56	}^{+	1.58	}$	\\

\hline \\[-0.5cm]
\label{table:orbitalParams3}
\end{tabular}
\end{minipage}
\end{table*}

\subsection{Comparison of the two methods}
The two methods used here to compute orbits are complementary, and we briefly comment here which method may be better suited for this work. The potential chosen in Galpy is fit to the observations of the MW, which works well for small to intermediate Galactic radii and beyond large radii (i.e. 50 kpc), the potential is less well known. The method using APOSTLE haloes, however, is less sensitive to the inner regions of the Galactic Centre, where multiple stellar components can influence the local potential. Therefore, when computing orbits within the inner regions of the Galaxy, a method such as Galpy might be more desirable. For this work we consider each method equally reliable for the MA8 halo trespasser due to the large orbital radius, however we believe MA11 and MA14 are better represented by orbits calculated with Galpy.

\section{Discussion}
\label{sec:disc}
\subsection{M22}
\label{sec:M22}

\subsubsection{CNO mixing}
\label{sec:m22CNO}
\begin{figure}
   \centering
    \includegraphics[clip=true,trim = 0 0 0 0,width=0.45\textwidth]{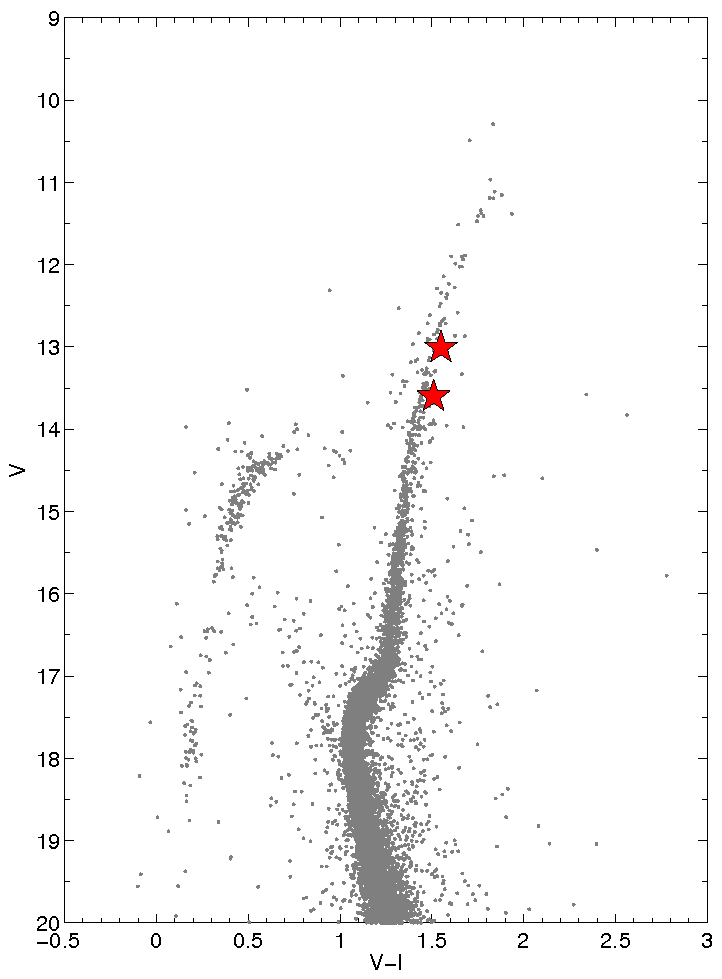}
   \caption{CMD of M22; photometry taken from the Hubble ACS Globular Cluster Survey \citep{Sarajedini2007}. Both target stars are above the RGB bump as indicated by their positions on the CMD (red stars).} 
   \label{fig:cmd}
\end{figure}

Stars above the RGB bump exhibit signs of mixing with CN-cycled material from the hydrogen burning shell, as has been seen in many globular clusters, e.g. \citet{SweigartMengel1979,SuntzeffSmith1991,Charbonnel1995,CharbonnelBrownWallerstein1998,Bellman2001,DenissenkovVandenBerg2003,SmithMartell2003,Carretta2005,SmithBriley2006}. Both of the stars in M22 analysed in this paper are above the RGB bump (see Figure \ref{fig:cmd}), and do show enhancements of N; however, this is also accompanied by enhancements in C and O. While both may be AGB stars (having also undergone the third dredge-up of gas from the H and He burning shells), it is also possible that the initial abundances of CNO are simply not scaled-solar.   The higher [O/Fe] value is consistent with halo stars at this metallicity, indicating initial abundances primarily from SNe II, unlike scaled-solar values which would include yields form SNeIa from a more extended galactic chemical evolution history.   If the initial C and N are also more pristine, then possibly [C/Fe]$_i$ $\sim$ [N/Fe]$_i$ $\sim$ +0.1 would be consistent then with mixing of CN-cycled gas; this is confirmed with CN-cycle calculator\footnote{http://www3.nd.edu/$\sim$vplacco/carbon-cor.html} as described in \citet{Placco2014}. \citet{Marino2011} and \citet{Alves-Brito2012} find two groups of stars in M22; one group that has A(C+N+O) $\sim$ 7.9 (and enhancements in s-process elements by [s/Fe]$\sim$0.35) and a second group that has A(C+N+O) $\sim$ 7.5 (and no s-process enrichment, thus [s/Fe]$\sim$0).
The two M22 stars in this paper have $<$A(C+N+O)$>$ = 7.46 ($\pm$ 0.25 for MA4, and $\pm$ 0.16 for MA4.1), consistent with the lower A(C+N+O) value. Note here the sum of C, N and O should be unaffected by the upper limit of C as discussed in Section \ref{sec:abAnal}.  This would imply that both are s-poor;  unfortunately, there are no s-process element spectral features in our H-band spectra.

 \subsubsection{Anti-correlations}

The Mg-Al anti-correlation in M22 may be evident in our stars, where we find enriched Mg and depleted Al relative to scaled-solar abundances. This anti-correlation was not evident in the work of \citet{Marino2011}. However, they do find a strong Na-O anti-correlation; this implies our M22 targets (with enhanced O abundances) are primordial, suggesting Al should be also be low since the Mg-Al cycle has not yet initiated. Indeed we find low Al (with enhanced O); this agrees with \citet{Marino2011}, who also find a strong Al-Na correlation, consequently implying an Al-O anti-correlation.Ó

\subsection{The bulge Candidates}

\subsubsection{CNO-cycling}
The physical gravities of our three bulge candidates would place them above the RGB bump on a stellar evolution track, and therefore they are likely to have undergone mixing with CN-cycled gas.   As discussed in Section \ref{sec:m22CNO} for our M22 targets, their high [O/Fe] implies that these stars do not simply have scaled-solar abundances for their metallicities, consistent with Galactic chemical evolution models. Similarly with MA8, we find elevated C and N values that imply that each of [(C,N,O)/Fe] are above scaled-solar values.     However, for MA11 and MA14, the elevated [N/Fe] and reduced [C/Fe] are consistent with scaled-solar CN-cycled gas (e.g., as found\footnote{The carbon correction yields $\Delta$[C/Fe]~=~+0.51, +0.65 for MA11 and MA14 (respectively).} with the interactive CN-cycling calculator as described in Section \ref{sec:m22CNO}).

\subsubsection{MA8: [Fe/H] = -1.9, r$_{apo}~>~$6.5 kpc, likely halo transient}

MA8 appears to be only $4.5^{+1.8}_{-1.3}$ kpc from the Sun towards the Galactic Centre; at $b$ = -5.2$^o$. A recent comprehensive review of the Galactic Centre distance by \citet{Bland-Hawthorn2016} considers 8.1 $\pm$ 0.1 kpc as the best distance estimate; MA8 is then 3.6 kpc from the Galactic Centre. Uncertainties in its proper motion result in orbits with a wide range in $r_\mathrm{{apo}}$ values as shown in Tables \ref{table:orbitalParams2} and \ref{table:orbitalParams3}, but all are larger than the size of the bulge.  This would imply that it is a halo star that is just passing through the Galactic bulge at this time.

The elemental abundances of MA8 are interesting when compared to halo stars, since the C, N, O, and Al abundances are slightly enhanced, while Mg and Si appear to be slightly depleted (see Figures \ref{fig:alphaCN} and \ref{fig:alphaA}). The Ni abundance is also unusual, although it has a large uncertainty. These unusual chemical abundances may suggest that it is associated with a disrupted dwarf galaxy, that had a unique chemical evolution history \citep[e.g.,][]{Tolstoy2009,Venn2004}, particularly when given the low [Mg/Ca] \citep[e.g.,][]{Lemasle2012,Venn2012, Jablonka2015}.

\subsubsection{MA11: [Fe/H] = -1.5, r$_{apo}~>~$3.2 kpc, possible bulge member}

The stellar parameters for MA11 imply that it is currently $11.0^{+4.3}_{-3.2}$ kpc from the Sun towards the Galactic Centre; at $b$ = 8.9$^o$, it would be directly above the Galactic Centre at a height of $\sim5$ kpc (consulting z$_{max}$ in Table \ref{table:orbitalParams3} computed with SPM4 proper motions);  given it's metallicity, it is possible MA11 may be a disc star, however this vertical orbital distance suggests this may not be the case. The lowest permitted apocenter value places MA11 in a bound orbit, however the majority of orbit calculations suggest it is a halo trespasser.

Its elemental abundances are consistent with halo, thick disk stars, and bulge stars in the literature, with the exception of perhaps Al. Al appears to be $\sim$2$\sigma$ higher than the halo stars, although there are very few halo stars at this metallicity (see Figure \ref{fig:alphaC}). Our Al is entirely consistent, however, with a bulge star from \cite{Johnson2013} at its metallicity. From these abundances and orbital kinematics, we argue that it is possible MA11 may be a bulge member.

\subsubsection{MA14: [Fe/H] = -2.1, r$_{apo}~>~$3.6 kpc, possible bulge member}

MA14 appears to be $4.5^{+1.9}_{-1.4}$ kpc from the Sun towards the Galactic Centre; at $b$ = -6.2$^o$; this places the object is 3.6 kpc from the Galactic Centre (again using the distance adopted in \citealt{Bland-Hawthorn}). As previously discussed in Section \ref{sec:orbitConclusion}, the orbit is sufficiently small that this could be a bulge member.

The elemental abundances for MA14 are consistent with those of halo stars, thick disk stars, and bulge stars at its respective metallicity, as indicated by Figures \ref{fig:alphaA}-\ref{fig:alphaC}. This is consistent with \citet{Melendez2008,Alves-Brito2010,Bensby2010,Bensby2011,Ryde2010,Gonzalez2011,Hill2011,Johnson2011,Johnson2013}), who suggest that the thick and metal-poor bulge populations have alpha-element abundances that are very similar and that the homogeneity between the two may indicate the bulge and the disk formed in situ. Based on these kinematic and elemental findings, we suggest that MA14 is a potential bulge member.

\section{Summary and Conclusions}

A detailed chemical abundance analysis has been performed on the H band IR spectra for metal-poor stars in and towards the Galactic Centre.  These spectra were taken with the newly developed technology of MOAO, using the RAVEN science demonstrator and IRCS detector at the Subaru 8.2-m telescope. In this analysis, we have found:

\begin{enumerate}
  \item The technical feasibility of MOAO on an 8-m telescope has been a success for high resolution spectroscopy. We report the first use of MOAO with high resolution spectroscopy, which is a crucial step in the science demonstration of this technology for the future ELT era. We demonstrate the benefits of MOAO and GLAO on the uncorrected PSF, yielding an on-average improvement of 0.2" in seeing, along with a 2-3 times flux gain through the spectroscopic slit. We also demonstrate these observations are successful when observing dusty, crowded regions within the MW.
  \item Five metal-poor stars in and towards the Galactic Centre have been identified and their chemical abundances were derived for 12 elements. Two of these stars belong to the globular cluster M22, while three of these stars are currently situated in the bulge.
  \item The two M22 stars have metallicities and element abundance ratios in common with other spectroscopic analyses \citep{Marino2011,Alves-Brito2012}.  Our CNO abundances are in common with the more metal-poor subpopulation in M22 and CNO-mixing, although we suggest the initial abundances may have been slightly higher than scaled-solar.
  \item The three metal-poor bulge candidates in this paper are in a metallicity regime with little to no previous high-resolution measurements (-1 $<$ [Fe/H] $<$ -2.5 dex).
  One out of three of our bulge candidates (MA8) has an orbit that suggests its a transiting halo object; its chemistry shows some unique abundance characteristics that could imply it was captured from an accreted dwarf galaxy.  Our two other stars (MA11 and MA14) may be bulge members based on both their kinematic and elemental properties. We note the large variation in orbit calculations when using two different proper motion catalogues. Future work on bulge members will benefit from the improved precision in the GAIA proper motions.
  \item The future of MOAO, and even GLAO, can benefit from the strategies developed for our observations, and from the lessons learned; we have summarized these strategies in order to facilitate the future use of this technology, which to date has had very little to no documented use.  We anticipate that in the coming ELT-era many interesting science cases will be developed specifically for the use of MOAO.
  \end{enumerate}

\vspace{10mm}
We are sincerely grateful to the RAVEN team, and the support of the Subaru staff during our engineering runs. MPL and KAV also acknowledge funding from an NSERC Discovery Grant to help fund this research.

\appendix
\section{Derived log abundances}
\begin{table*}
	\centering
	\caption{Atomic lines and derived log abundances}

			\begin{tabular}{@{}lccccccccc@{}}
			\hline
Element	&	Lambda ($\mathrm{\AA}$)	&	$\chi$ (eV)	&	log $gf$	&	M22-MA4	&	M22MA4.1	&	MA8	&	MA11	&	MA14	&	M15K341	\\
\hline
FeI	&	15051.75	&	5.35	&	0.21	&	-	&	5.83	&	-	&	6.01	&	5.68	&	-	\\
	&	15194.49	&	2.22	&	-4.76	&	-	&	-	&	-	&	-	&	5.58	&	-	\\
	&	15207.53	&	5.39	&	0.17	&	5.32	&	5.43	&	-	&	5.91	&	5.38	&	4.97	\\
	&	15219.62	&	5.59	&	-0.03	&	5.47	&	5.53	&	-	&	6.01	&	5.18	&	-	\\
	&	15294.56	&	5.31	&	0.75	&	5.17	&	5.28	&	5.23	&	5.61	&	5.28	&	4.92	\\
	&	15394.67	&	5.62	&	-0.23	&	-	&	5.48	&	-	&	5.91	&	5.38	&	-	\\
	&	15395.72	&	5.62	&	-0.30	&	-	&	-	&	-	&	6.01	&	5.58	&	-	\\
	&	15531.75	&	5.64	&	-0.54	&	-	&	-	&	-	&	-	&	5.58	&	-	\\
	&	15588.26	&	6.37	&	0.32	&	-	&	-	&	-	&	5.91	&	-	&	-	\\
	&	15591.50	&	6.36	&	0.95	&	5.37	&	5.63	&	5.53	&	6.06	&	5.33	&	-	\\
	&	15604.22	&	6.24	&	0.44	&	5.52	&	5.68	&	-	&	6.06	&	5.23	&		\\
	&	15621.65	&	5.54	&	0.35	&	5.27	&	5.53	&	-	&	5.91	&	5.38	&	5.12	\\
	&	15631.95	&	5.35	&	-0.02	&	5.42	&	5.48	&	-	&	6.16	&	5.38	&	5.42	\\
	&	15648.51	&	5.43	&	-0.66	&	-	&	5.68	&	5.93	&	6.01	&	5.38	&	-	\\
	&	15662.02	&	5.83	&	0.18	&	-	&	-	&	-	&	5.91	&	-	&	5.32	\\
	&	15677.52	&	6.25	&	0.21	&	-	&	5.78	&	-	&	6.01	&	5.63	&	-	\\
	&	15686.44	&	6.25	&	0.11	&	-	&	5.73	&	-	&	6.06	&	5.53	&	-	\\
	&	15691.86	&	6.25	&	0.47	&	5.67	&	5.58	&	-	&	5.96	&	5.38	&	5.47	\\
	&	15723.59	&	5.62	&	0.03	&	-	&	5.63	&	-	&	5.96	&	5.28	&	-	\\
	&	15741.92	&	5.65	&	-0.40	&	-	&	-	&	-	&	6.11	&	5.58	&	5.77	\\
	&	15769.42	&	5.54	&	-0.01	&	5.72	&	5.83	&	-	&	6.16	&	5.58	&	5.42	\\
	&	15789.00	&	6.25	&	0.32	&	5.82	&	-	&	-	&	6.11	&	-	&	-	\\
	&	15798.56	&	6.25	&	0.34	&	-	&	5.78	&	-	&	6.16	&	-	&	-	\\
	&	15818.14	&	5.59	&	0.32	&	5.52	&	5.63	&	5.33	&	5.96	&	5.28	&	5.27	\\
	&	15822.82	&	5.64	&	-0.10	&	5.67	&	5.73	&	5.83	&	5.96	&	-	&	-	\\
	&	15835.17	&	6.30	&	0.67	&	-	&	5.58	&	5.33	&	5.86	&	5.68	&	5.57	\\
	&	15837.65	&	6.30	&	0.16	&	-	&	5.98	&	-	&	5.96	&	5.78	&	-	\\
	&	15868.52	&	5.59	&	-0.25	&	-	&	5.33	&	-	&	5.91	&	5.08	&	5.22	\\
	&	15878.44	&	5.62	&	-0.48	&	-	&	-	&	-	&	-	&	5.83	&	-	\\
	&	15895.23	&	6.26	&	0.28	&	-	&	-	&	-	&	6.16	&	-	&	-	\\
	&	15906.04	&	5.62	&	-0.16	&	-	&	5.43	&	5.43	&	-	&	5.48	&	-	\\
	&	15980.73	&	6.26	&	1.12	&	4.92	&	5.43	&	-	&	5.71	&	-	&	-	\\
	&	16006.76	&	6.35	&	0.63	&	-	&	-	&	-	&	5.91	&	-	&	-	\\
	&	16009.61	&	5.43	&	-0.53	&	5.87	&	5.58	&	5.38	&	6.06	&	-	&	5.52	\\
	&	16040.66	&	5.87	&	0.12	&	-	&	5.63	&	-	&	6.11	&	-	&	-	\\
	&	16042.71	&	6.26	&	0.02	&	-	&	-	&	-	&	6.01	&	-	&	-	\\
	&	16102.41	&	5.87	&	0.26	&	5.47	&	-	&	5.68	&	5.96	&	5.58	&	-	\\
	&	16115.97	&	6.39	&	0.32	&	-	&	-	&	5.83	&	5.96	&	-	&	-	\\
	&	16125.90	&	6.35	&	0.69	&	5.27	&	5.73	&	5.23	&	5.91	&	5.73	&	5.32	\\
	&	16153.25	&	5.35	&	-0.70	&	-	&	5.58	&	5.68	&	6.01	&	-	&	5.52	\\
	&	16165.03	&	6.32	&	0.79	&	5.62	&	5.73	&	6.03	&	6.16	&	5.48	&	-	\\
	&	16174.95	&	6.38	&	0.01	&	-	&	-	&	-	&	6.06	&	5.73	&	-	\\
	&	16198.50	&	5.41	&	-0.49	&	5.47	&	5.63	&	-	&	5.81	&	-	&	-	\\
	&	16231.65	&	6.38	&	0.51	&	5.57	&	-	&	-	&	-	&	-	&	-	\\
	&	16316.32	&	6.28	&	1.12	&	5.42	&	5.23	&	5.78	&	5.86	&	5.08	&	4.92	\\
	&	16324.45	&	5.39	&	-0.55	&	-	&	5.53	&	-	&	5.86	&	5.28	&	5.42	\\
	&	16444.82	&	5.83	&	0.41	&	-	&	-	&	-	&	6.26	&	-	&	-	\\
	&	16486.67	&	5.83	&	0.52	&	5.62	&	5.78	&	5.53	&	5.91	&	5.48	&	5.25	\\
	&	16506.30	&	5.95	&	-0.37	&	-	&	-	&	-	&	-	&	5.53	&	-	\\
	&	16517.23	&	6.29	&	0.56	&	5.62	&	5.38	&	5.53	&	6.11	&	5.43	&	-	\\
	&	16524.47	&	6.34	&	0.64	&	-	&	5.63	&	-	&	6.06	&	-	&	-	\\
	&	16541.43	&	5.95	&	-0.40	&	-	&	-	&	-	&	5.96	&	5.58	&	-	\\
	&	16552.00	&	6.41	&	0.12	&	-	&	-	&	-	&	6.26	&	-	&	-	\\
	&	16561.77	&	5.98	&	0.11	&	5.47	&	5.53	&	5.33	&	6.11	&	5.28	&	-	\\
	&	16969.91	&	5.95	&	-0.20	&	-	&	-	&	-	&	5.81	&	5.59	&	-	\\

				\hline
			\end{tabular}
			\label{table:bla}
\end{table*}	

\begin{table*}
	\centering
	\caption{Atomic lines and derived log abundances - continued}

			\begin{tabular}{@{}lccccccccc@{}}
			\hline
Element	&	Lambda ($\mathrm{\AA}$)	&	$\chi$ (eV)	&	log $gf$	&	M22-MA4	&	M22MA4.1	&	MA8	&	MA11	&	MA14	&	M15K341	\\
\hline

Mg	&	15024.99	&	5.11	&	0.37	&	-	&	5.77	&	-	&	6.55	&	5.90	&	-	\\
	&	15040.25	&	5.11	&	0.13	&	-	&	5.92	&	-	&	6.55	&	6.00	&	-	\\
	&	15047.71	&	5.11	&	-0.37	&	-	&	5.77	&	-	&	6.30	&	5.90	&	-	\\
	&	15740.72	&	5.93	&	-0.26	&	6.05	&	5.82	&	5.75	&	6.40	&	5.80	&	5.47	\\
	&	15748.99	&	5.93	&	0.17	&	5.80	&	5.87	&	6.00	&	6.55	&	5.70	&	5.57	\\
	&	15765.84	&	5.93	&	0.44	&	5.75	&	5.82	&	5.85	&	6.55	&	5.65	&	5.52	\\
	&	15879.57	&	5.95	&	-1.30	&	6.25	&	-	&	-	&	6.45	&	-	&	-	\\
	&	15886.18	&	5.95	&	-1.65	&	-	&	6.17	&	-	&	6.75	&	-	&	-	\\
	&	16624.72	&	6.73	&	-1.48	&	-	&	-	&	-	&	6.85	&	-	&	-	\\
	&	16632.02	&	6.73	&	-1.31	&	-	&	-	&	-	&	6.85	&	-	&		\\
	&		&		&		&		&		&		&		&		&		\\
Al	&	16718.96	&	4.09	&	0.29	&	4.04	&	4.26	&	4.49	&	5.14	&	3.99	&	4.06	\\
	&	16750.56	&	4.09	&	0.55	&	4.24	&	4.21	&	4.24	&	5.24	&	4.09	&	3.96	\\
	&	16763.36	&	4.09	&	-0.53	&	-	&	4.56	&	-	&	5.14	&	-	&	-	\\
	&		&		&		&		&		&		&		&		&		\\
Si	&	15376.83	&	6.22	&	-0.58	&	6.07	&	5.99	&	-	&	6.28	&	5.87	&	-	\\
	&	15557.78	&	5.96	&	-0.68	&	6.07	&	5.74	&	5.87	&	6.18	&	5.77	&	5.54	\\
	&	15833.60	&	6.22	&	-0.17	&	6.22	&	5.69	&	5.62	&	6.20	&	5.52	&	5.24	\\
	&	15884.45	&	5.95	&	-0.69	&	5.52	&	5.69	&	5.72	&	6.15	&	-	&	5.54	\\
	&	15888.41	&	5.08	&	0.02	&	6.17	&	5.69	&	5.72	&	6.20	&	5.77	&	5.44	\\
	&	16060.01	&	5.95	&	-0.48	&	5.82	&	5.79	&	5.92	&	6.25	&	5.77	&	5.44	\\
	&	16094.79	&	5.96	&	-0.16	&	5.52	&	5.84	&	5.62	&	6.35	&	5.67	&	5.54	\\
	&	16163.69	&	5.95	&	-0.94	&	5.82	&	5.94	&	5.82	&	6.45	&	5.87	&	5.64	\\
	&	16215.67	&	5.95	&	-0.60	&	5.52	&	5.84	&	5.62	&	6.30	&	5.77	&	5.24	\\
	&	16241.83	&	5.96	&	-0.77	&	5.82	&	5.79	&	5.52	&	6.30	&	5.72	&	5.24	\\
	&	16434.93	&	5.96	&	-1.15	&		&	-	&	-	&	6.15	&	-	&	-	\\
	&	16680.77	&	5.98	&	-0.06	&	5.12	&	5.74	&	5.72	&	6.35	&	5.67	&	5.24	\\
	&	16828.16	&	5.98	&	-1.08	&	5.52	&	5.74	&	-	&	6.65	&	5.87	&	5.64	\\
	&		&		&		&		&		&		&		&		&		\\
S	&	15403.72	&	8.70	&	-0.14	&	5.73	&	5.80	&	5.73	&	5.93	&	-	&	-	\\
	&	15469.82	&	8.05	&	-0.15	&	6.08	&	-	&	5.88	&	6.13	&	-	&	-	\\
	&	15478.48	&	8.05	&	0.08	&	5.48	&	5.60	&	-	&	-	&	-	&	-	\\
	&		&		&		&		&		&		&		&		&		\\
Ca	&	16150.76	&	4.53	&	-0.17	&	-	&	4.55	&	4.38	&	5.08	&	4.43	&	-	\\
	&	16155.24	&	4.53	&	-0.58	&	4.53	&	-	&	4.98	&	5.18	&	4.88	&	-	\\
	&	16157.36	&	4.55	&	-0.14	&	4.63	&	4.65	&	4.78	&	5.13	&	-	&	-	\\
	&	16197.08	&	4.54	&	0.16	&	4.63	&	4.15	&	-	&	5.08	&	4.48	&	-	\\
	&		&		&		&		&		&		&		&		&		\\
Ti	&	15334.85	&	1.89	&	-1.00	&	3.61	&	3.48	&	3.11	&	-	&	-	&	-	\\
	&	15543.76	&	1.88	&	-1.12	&	3.56	&	2.98	&	3.41	&	$<$3.66	&	$<$3.16	&	-	\\
	&		&		&		&		&		&		&		&		&		\\
Mn	&	15159.14	&	4.89	&	-0.06	&	$<$3.66	&	$<$3.48	&	-	&	3.86	&	$<$3.31	&	$<$3.08	\\
	&	15217.76	&	4.89	&	-0.19	&	-	&	-	&	-	&	3.86	&	-	&	-	\\
	&	15262.50	&	4.89	&	-0.27	&	-	&	-	&	-	&	4.11	&	-	&	-	\\
	&		&		&		&		&		&		&		&		&		\\
Ni	&	15555.42	&	5.49	&	0.10	&	4.47	&	4.64	&	5.02	&	-	&	-	&	-	\\
	&	15605.68	&	5.30	&	-0.45	&	4.77	&	-	&	-	&	4.92	&	4.47	&	-	\\
	&	15632.65	&	5.31	&	-0.04	&	4.22	&	-	&	-	&	4.72	&	4.17	&	-	\\
	&	16136.16	&	5.49	&	-0.14	&	-	&	-	&	-	&	4.62	&	-	&	-	\\
	&	16310.50	&	5.28	&	0.00	&	4.52	&	4.24	&	4.12	&	4.22	&	3.87	&	3.91	\\
	&	16363.11	&	5.28	&	0.37	&	4.52	&	-	&	-	&	-	&	3.97	&	3.41	\\
	&	16589.29	&	5.47	&	-0.55	&	-	&	-	&	-	&	-	&	-	&	4.51	\\
	&	16673.71	&	6.03	&	0.39	&	-	&	-	&	-	&	4.92	&	-	&	-	\\
	&	16996.27	&	5.30	&	0.34	&	-	&	4.29	&	-	&	4.72	&	4.17	&	-	\\
				\hline
			\end{tabular}
			\label{table:bla}
\end{table*}	

\begin{table*}
	\centering
	\caption{Molecular features used to derive C, N, and O and their log abundances}

			\begin{tabular}{@{}lccccccc@{}}
			\hline
Element		&	Lambda interval ($\mathrm{\AA}$)	&	M22-MA4	&	M22MA4.1	&	MA8	&	MA11	&	MA14	&	M15K341	\\
\hline
C from CO lines	&	15582-15590	&	6.93	&	6.75	&	-	&	6.73	&	6.34	&	6.15	\\
	&	15780-15788	&	-	&	6.85	&	6.68	&	6.93	&	5.94	&	6.15	\\
	&	15977-15997	&	-	&	6.80	&	-	&	6.33	&	-	&	6.27	\\
	&	16000-16018	&	5.98	&	-	&	-	&	6.53	&	5.89	&	-	\\
	&	16182-16189	&	6.37	&	6.60	&	-	&	6.33	&	-	&	5.82	\\
	&	16217-16229	&	-	&	-	&	-	&	-	&	-	&	-	\\
	&	16304-16309	&	-	&	-	&	7.03	&	-	&	-	&	-	\\
	&	16475-16485	&	-	&	-	&	6.68	&	-	&	6.14	&	-	\\
	&	16609-16649	&	-	&	6.65	&	6.73	&	6.63	&	-	&	-	\\
	&	16667-16679	&	-	&	6.95	&	7.23	&	-	&	-	&	-	\\
	&	16690-16702	&	-	&	-	&	6.98	&	-	&	5.79	&	-	\\
	&	16836-16851	&	-	&	-	&	6.63	&	-	&	-	&	-	\\
	&		&		&		&		&		&		&		\\
O from OH lines	&	15002	&	7.50	&	7.32	&	-	&	7.80	&	7.55	&	-	\\
	&	15004	&	-	&	-	&	-	&	7.40	&	7.45	&	-	\\
	&	15021	&	-	&	7.72	&	-	&	7.55	&	7.45	&	-	\\
	&	15023	&	-	&	-	&	-	&	7.40	&	7.40	&	-	\\
	&	15129	&	-	&	7.02	&	-	&	-	&	7.30	&	6.82	\\
	&	15131	&	-	&	-	&	-	&	7.40	&	7.40	&	7.07	\\
	&	15145	&	-	&	7.32	&	-	&	7.50	&	7.35	&	6.82	\\
	&	15148	&	-	&	-	&	-	&	-	&	-	&	6.87	\\
	&	15237	&	-	&	-	&	-	&	7.30	&	7.15	&	-	\\
	&	15266	&	-	&	7.22	&	7.50	&	7.20	&	7.30	&	6.97	\\
	&	15279	&	7.60	&	7.17	&	7.90	&	-	&	7.35	&	7.02	\\
	&	15281	&	7.35	&	-	&	7.60	&	-	&	7.30	&	7.02	\\
	&	15392	&	-	&	-	&	7.75	&	7.10	&	7.25	&	6.72	\\
	&	15409	&	-	&	7.32	&	7.55	&	-	&	7.30	&	6.92	\\
	&	15422	&	-	&	7.22	&	7.45	&	7.55	&	7.35	&	6.97	\\
	&	15560	&	-	&	7.42	&	7.55	&	-	&	-	&	6.92	\\
	&	15569	&	-	&	7.32	&	7.50	&	-	&	7.30	&	6.82	\\
	&	15572	&	-	&	7.22	&	-	&	7.50	&	7.35	&	7.07	\\
	&	15627	&	-	&	7.02	&	7.40	&	-	&	7.30	&	7.09	\\
	&	15652	&	-	&	7.42	&	7.40	&	7.50	&	7.30	&	7.07	\\
	&	15654	&	-	&	7.42	&	7.50	&	7.45	&	7.35	&	7.07	\\
	&	15719	&	-	&	7.22	&	7.65	&	7.45	&	7.25	&	-	\\
	&	15726	&	7.50	&	-	&	7.90	&	7.40	&	7.30	&	7.02	\\
	&	15755	&	-	&	-	&	7.70	&	-	&	-	&	6.92	\\
	&	15756	&	7.20	&	7.37	&	-	&	7.60	&	7.25	&	6.62	\\
	&	15893	&	-	&	-	&	7.40	&	-	&	7.35	&	-	\\
	&	15911	&	7.20	&	7.12	&	7.85	&	7.60	&	7.35	&	6.97	\\
	&	16037	&	-	&	7.42	&	-	&	7.40	&	7.15	&	6.87	\\
	&	16039	&	-	&	7.22	&	7.70	&	-	&	-	&	6.97	\\
	&	16052	&	-	&	7.12	&	7.80	&	7.75	&	7.30	&	6.92	\\
	&	16061	&	-	&	7.12	&	7.65	&	-	&	-	&	6.77	\\
	&	16065	&	-	&	7.27	&	7.60	&	-	&	-	&	-	\\
	&	16069	&	-	&	6.92	&	7.35	&	-	&	-	&	6.92	\\
	&	16074	&	-	&	7.22	&	7.50	&	-	&	-	&	-	\\
	&	16190	&	-	&	-	&	-	&	-	&	7.15	&	-	\\
	&	16192	&	-	&	-	&	-	&	-	&	7.05	&	-	\\
	&	16248	&	-	&	-	&	-	&	-	&	-	&	6.92	\\
	&	16252	&	-	&	7.22	&	7.45	&	-	&	-	&	6.97	\\
	&	16255	&	-	&	7.12	&	7.65	&	-	&	-	&	6.92	\\
	&	16260	&	-	&	7.22	&	7.00	&	-	&	-	&	6.92	\\
	&	16313	&	7.10	&	7.37	&	7.70	&	7.55	&	7.15	&	-	\\
	&	16346	&	-	&	7.72	&	7.50	&	-	&	-	&	-	\\
	&	16348	&	-	&	7.02	&	-	&	-	&	-	&	-	\\
	&	16352	&	-	&	7.12	&	7.75	&	-	&	-	&	6.72	\\
	&	16354	&	-	&	7.42	&	7.35	&	-	&	-	&	6.97	\\
	&	16368	&	-	&	-	&	-	&	-	&	7.20	&	-	\\
	&	16448	&	-	&	7.37	&	-	&	7.60	&	7.25	&	-	\\
	&	16450	&	7.20	&	7.32	&	-	&	7.75	&	7.10	&	-	\\
	&	16456	&	-	&	7.37	&	-	&	6.90	&	7.30	&	-	\\
	&	16523	&	-	&	7.17	&	-	&	7.40	&	7.25	&	-	\\
				\hline
			\end{tabular}
			\label{table:bla2}
\end{table*}	

\begin{table*}
	\centering
	\caption{Molecular features used to derive C, N, and O - continued}

			\begin{tabular}{@{}lccccccc@{}}
			\hline
Molecular lines		&	Lambda interval ($\mathrm{\AA}$)	&	M22-MA4	&	M22MA4.1	&	MA8	&	MA11	&	MA14	&	M15K341	\\
\hline
O from OH lines	&	16526	&	-	&	-	&	7.45	&	7.45	&	7.20	&	6.87	\\
	&	16535	&	7.30	&	7.27	&	-	&	7.15	&	7.30	&	6.97	\\
	&	16539	&	7.50	&		&	7.70	&	7.45	&	7.05	&	6.92	\\
	&	16605	&	-	&	7.12	&	7.60	&	7.70	&	7.15	&	7.22	\\
	&	16608	&	-	&	-	&	7.70	&	-	&	-	&	7.22	\\
	&	16704	&	-	&	-	&	7.55	&	7.40	&	7.15	&	6.82	\\
	&	16708	&	7.20	&	-	&	7.75	&	7.10	&	7.05	&	6.77	\\
	&	16714	&	-	&	-	&	7.80	&	-	&	-	&	6.82	\\
	&		&		&		&		&		&		&		\\
N from CN lines	&	15004	&	-	&	6.04	&	-	&	-	&	6.57	&	-	\\
	&	15012	&	-	&	6.29	&	-	&	6.92	&	-	&	-	\\
	&	15034-15036	&	-	&	-	&	-	&	-	&	6.87	&	-	\\
	&	15048	&	-	&	-	&	-	&	6.92	&	-	&	-	\\
	&	15052	&	-	&	-	&	-	&	7.02	&	-	&	-	\\
	&	15106	&	-	&	-	&	-	&	7.02	&	-	&	-	\\
	&	15118	&	-	&	-	&	-	&	6.87	&	6.67	&	-	\\
	&	15128	&	-	&	-	&	-	&	-	&	7.07	&	-	\\
	&	15134-15139	&	-	&	6.74	&	-	&	-	&	-	&	6.34	\\
	&	15142	&	-	&	-	&	-	&	-	&	-	&	6.34	\\
	&	15150-15166	&	-	&	6.34	&	-	&	-	&	6.67	&	6.34	\\
	&	15184-15200	&	-	&	6.19	&	-	&	6.82	&	6.97	&	6.14	\\
	&	15210	&	-	&	-	&	-	&	6.82	&	6.97	&	-	\\
	&	15222	&	-	&	-	&	-	&	-	&	6.87	&	-	\\
	&	15228	&	-	&	-	&	-	&	-	&	6.77	&	-	\\
	&	15242	&	6.88	&	6.19	&	-	&	-	&	-	&	-	\\
	&	15251-15254	&	-	&	6.19	&	-	&	6.77	&	6.62	&	6.29	\\
	&	15284-15287	&	-	&	-	&	6.62	&	-	&	-	&	-	\\
	&	15308-15323	&	6.23	&	-	&	6.27	&	6.92	&	6.82	&	6.09	\\
	&	15328	&	-	&	-	&	-	&	-	&	-	&	6.09	\\
	&	15351-15362	&	6.53	&	6.04	&	6.37	&	6.82	&	6.77	&	-	\\
	&	15363	&	-	&	-	&	6.87	&	-	&	-	&	-	\\
	&	15389	&	-	&	-	&	6.17	&	6.62	&	-	&	-	\\
	&	15400	&	-	&	-	&	6.62	&	-	&	-	&	-	\\
	&	15411	&	-	&	6.19	&	-	&	-	&	-	&	6.39	\\
	&	15432	&	-	&	-	&	-	&	7.02	&	-	&	-	\\
	&	15448	&	-	&	6.54	&	-	&	-	&	-	&	-	\\
	&	15472	&	-	&	6.39	&	-	&	-	&	-	&	-	\\
	&	15564	&	7.13	&	6.59	&	-	&	-	&	-	&	-	\\
	&	15575	&	-	&	-	&	-	&	6.82	&	-	&	-	\\
	&	15595	&	-	&	6.84	&	-	&	-	&	6.72	&	-	\\
	&	15636	&	-	&	6.19	&	-	&	-	&	6.77	&	-	\\
	&	15660	&	-	&	6.39	&	-	&	7.12	&	-	&	-	\\
	&	15767-15775	&	-	&	-	&	6.12	&	-	&	-	&	-	\\
	&	15825	&	-	&	-	&	7.02	&	-	&	-	&	-	\\
	&	16180	&	-	&	-	&	6.52	&	-	&	-	&	-	\\
				\hline
			\end{tabular}
			\label{table:bla2}
\end{table*}	

\label{lastpage}
\end{document}